\documentclass[aps,prd,amsmath,preprint]{revtex4-1}

\usepackage[utf8]{inputenc}
\usepackage{enumerate}
\usepackage{amsmath,amssymb}
\usepackage{amsfonts}
\usepackage{bm}
\usepackage{amscd}            
\usepackage{graphicx}
\usepackage{hhline,multirow}  
\usepackage{dcolumn}          
\usepackage{slashed}
\usepackage{braket}
\usepackage{url}              
\usepackage[colorlinks=true,linkcolor=blue]{hyperref}
\usepackage{color}
\usepackage{soul}
\usepackage{float}

\preprint{JLAB-THY-15-2183}

\begin{document}

\title{Pion structure function from leading neutron electroproduction
	and SU(2) flavor asymmetry}

\author{J.~R.~McKenney$^{1,2,3}$,
	Nobuo~Sato$^2$,
	W.~Melnitchouk$^2$,
	Chueng-Ryong~Ji$^1$}
\affiliation{
$^1$\mbox{North Carolina State University,
	  Raleigh, North Carolina 27695, USA} \\
$^2$\mbox{Jefferson Lab, Newport News,
	  Virginia 23606, USA} \\
$^3$\mbox{University of North Carolina,
	  Chapel Hill, North Carolina 27599, USA} \\
}

\date{\today}

\begin{abstract}
We examine the efficacy of pion exchange models to simultaneously
describe leading neutron electroproduction at HERA and the
$\bar{d}-\bar{u}$ flavor asymmetry in the proton.
A detailed $\chi^2$ analysis of the ZEUS and H1 cross sections,
when combined with constraints on the pion flux from Drell-Yan
data, allows regions of applicability of one-pion exchange to be
delineated.
The analysis disfavors several models of the pion flux used in the
literature, and yields an improved extraction of the pion structure
function and its uncertainties at parton momentum fractions in the
pion of $4 \times 10^{-4} \lesssim x_\pi \lesssim 0.05$ at a scale
of $Q^2=10$~GeV$^2$.
Based on the fit results, we provide estimates for leading proton
structure functions in upcoming tagged deep-inelastic scattering
experiments at Jefferson Lab on the deuteron with forward protons.
\end{abstract}

\maketitle

\section{Introduction}
\label{sec:intro}

The importance of pions in the structure and interactions of nucleons
has been known since the discoveries of the neutron in the 1930s
\cite{Chadwick32} and of the pion itself in the 1940s \cite{Powell47,
Yukawa35}.  Long recognized to be the bosonic mediators of the
long-range part of the nucleon--nucleon force, the role of pions in
nuclear interactions has in recent decades been codified in the form
of chiral effective theory, exploiting the approximate chiral symmetry
properties of the fundamental QCD lagrangian.

Despite the tremendous progress made in understanding the consequences
of chiral symmetry breaking for nuclear and hadron phenomenology
\cite{Thomas84, Machleidt01, Meissner10}, many aspects of pion physics
still remain elusive.
Indeed, the pion presents itself as a dichotomy, with its simultaneous
existence as the pseudo-Goldstone boson associated with chiral symmetry
breaking in QCD, and as the lightest QCD bound state composed of quark
and gluon (or parton) constituents \cite{WMdual}.
The partonic nature of the pion is revealed most clearly in
high-energy processes, which are most efficiently formulated on the
light-front; on the other hand, the description of low-energy chiral
physics on the light-front has historically been challenging and
remains an important area of modern research \cite{Beane13, Granados15}.

From the purely phenomenological perspective, study of the consequences
of chiral symmetry breaking and the role of the pion has provided many
insights into the structure of the nucleon, from the electromagnetic
charge distribution of the neutron to the nuclear EMC effect and the
modification of nucleon properties in the nuclear medium.
One of the most dramatic consequences of the nucleon's pion cloud has
been in the flavor structure of the proton sea, with the finding of
a large excess of $\bar d$ quarks in the proton over $\bar u$.
First anticipated by Thomas \cite{Thomas83} in the 1980s on the basis
of the scaling properties of one-pion exchange in deep-inelastic
scattering (DIS) \cite{Sullivan72}, the empirical observation of
a large $\bar d-\bar u$ asymmetry by the New Muon Collaboration
\cite{NMC} at CERN, and later even more conclusively by the E866
Collaboration \cite{E866} at Fermilab, firmly established the
relevance of pions for understanding the partonic structure of
the nucleon \cite{Speth98, Garvey01}.

In the subsequent years much successful phenomenology has been
developed in applying pion cloud models to the nucleon's
nonperturbative structure, although the connection with the
underlying QCD theory has not always been manifest.
The difficulty reflects the question of how to apply effective
chiral theory techniques, which are formally grounded in the
symmetries of QCD, to observables accessible at high energies,
where the degrees of freedom are not those of the effective theories.
Recently, however, progress in linking pionic effects in partonic
observables directly with QCD has been made by considering the
nonanalytic structure of matrix elements expanded in terms of
the pion mass, $m_\pi$.
In particular, in analogy with low-energy observables such as masses
and magnetic moments, it was found that moments of parton distribution
functions (PDFs) could be systematically expanded in powers of
$m_\pi^2$, with the coefficients of the leading nonanalytic (LNA)
terms given in terms of model-independent constants \cite{TMS00,
Detmold01, Chen01, Arndt02, Dorati07}.
This enabled an unambiguous connection to be established between
chiral symmetry breaking in QCD and the existence of an SU(2)
flavor asymmetry in the proton \cite{TMS00}.

Building on these earlier observations, more recent studies have
sought to develop the phenomenology of nonperturbative parton
distributions in the context of chiral effective theory, not just
in terms of moments but also as a function of the parton momentum
fraction $x$ \cite{Burkardt13, Salamu15}.
While much of the attention has been focused on exploring the
consequences of chiral symmetry breaking for the $\bar d-\bar u$
asymmetry in the proton, widely seen as the ``smoking gun'' signal
of the pion cloud, a complementary effort to reveal the dynamics
of pion exchange in high-energy processes has been the study of
leading neutron production in semi-inclusive DIS on the proton.
Here a forward moving neutron is produced in coincidence with the
scattered lepton in the high-energy reaction $e p \to e n X$,
and several dedicated experiments at the $ep$ collider HERA
\cite{ZEUS_02, H1_10, ZEUS_07} have collected high-precision data
on the spectrum of leading neutrons carrying a large fraction of
the proton's energy.

As well as identifying the characteristic features of pion exchange
in the leading neutron production cross sections, the HERA data have
also been analyzed in view of extracting the structure function of
the exchanged pion in the small-$x_\pi$ region \cite{ZEUS_02, H1_10,
Holtmann94, D'Alesio00, Kopeliovich12, Carvalho15}.
Previous determinations of the PDFs in the pion based on fits to
Drell-Yan and prompt photon production data from $\pi N$ scattering
experiments at CERN \cite{NA3, NA10} and Fermilab \cite{E615} have
typically been restricted to the high-$x_\pi$ region
($x_\pi \gtrsim 0.2$).  Analyses of the HERA
leading neutron data have generally been able to extract the shape of
the pion structure function $F_2^\pi$, but have been unable to fix
the normalization because of large uncertainties in the pion flux
(or pion light-cone momentum distribution in the nucleon).
Since the pionic contributions to the leading neutron cross sections
depend on both the pion structure function and the pion probability
in the proton, the HERA data by themselves have been insufficient to
disentangle information on $F_2^\pi$ independently of assumptions
about the pion flux.

On the other hand, a systematic study of the assumptions about
the pion distribution function has not yet been performed.
The ZEUS analysis of their data \cite{ZEUS_02} used as a baseline
a Regge theory inspired model of the pion flux \cite{Bishari},
but found a factor 2 difference in the normalization of $F_2^\pi$
when compared with an additive quark model.
Earlier, D'Alesio \& Pirner \cite{D'Alesio00} considered models of
the pion distribution function in $pp$ scattering using a traditional
$t$-dependent $\pi NN$ form factor, as well as a light-cone inspired
form, with parameters fixed from inclusive neutron production data.
Because the absorptive corrections in $pp$ versus $\gamma^* p$
scattering are expected to be different, however, it was argued
\cite{D'Alesio00} that this jeopardized the possibility of a
reliable extraction of $F_2^\pi$ to be made.

More recently, Kopeliovich {\it et al.} \cite{Kopeliovich12} used
a Reggeized pion exchange model, supplemented by vector and axial
vector mesons and absorption corrections, to study leading neutron
spectra within a dipole approach.
Assuming the ratio of the pion to proton structure functions
to be proportional to the ratio of the number of quarks in the
respective hadrons, $N_q^\pi/N_q^p$, the comparison with the HERA
data suggested the extracted $F_2^\pi$ would be somewhat sensitive
to the precise value of $N_q^\pi/N_q^p$, as well as to the coherence
length parametrizing the absorptive corrections.
The color dipole model for the virtual photon--pion cross section
was also used recently by Carvalho {\it et al.} \cite{Carvalho15}
to study gluon saturation effects at small $x$, using a range of
$\pi NN$ form factor models from the literature.
In an alternative approach, de~Florian and Sassot \cite{deFlorian97}
formulated the one-pion exchange contributions to the leading neutron
cross section in terms of fracture functions.  While the fracture
functions are more general constructs, in the pion model they can be
computed as products of the pion flux and pion structure function.

In the present analysis we wish to address the question of whether
one can reduce the model dependence of $F_2^\pi$ extracted from the
HERA leading neutron data by using additional constraints from other
observables that are sensitive to the pion flux.
In particular, the data on the SU(2) flavor asymmetry $\bar d-\bar u$,
particularly those from the E866 Drell-Yan experiment \cite{E866},
provide the strongest indication of significant pion cloud effects
in the nucleon.
Because the E866 data are at relatively high $x$ values compared with
the HERA measurements, within the pion exchange framework they are
sensitive to the pion PDFs at large $x_\pi$, where the PDFs are well
determined from pion--nucleon Drell-Yan data \cite{NA3, NA10, E615}.
The main variable in describing the $\bar d-\bar u$ asymmetry is
therefore the pion distribution function in the nucleon.

In contrast, the HERA data are taken at very low $x$,
$10^{-4} \lesssim x \lesssim 10^{-2}$, outside of the region where
the pion PDFs have been constrained.  Within the pion exchange
framework, the same pion flux should be applicable for both
observables, which should then reduce the uncertainty in the
extracted $F_2^\pi$ at small $x$.
Surprisingly, a quantitative analysis of this type has never been
performed.  In this study we use methodology adopted from global
PDF analysis \cite{Jimenez13, Forte13} to simultaneously fit both
the HERA leading neutron and E866 $\bar d-\bar u$ asymmetry data.

In Sec.~\ref{sec:pion} we begin by reviewing pion exchange models,
summarizing the main results for pion distribution functions in the
nucleon derived from chiral effective theory, and discussing various
regularization prescriptions that have been used in the literature
for the hadronic $\pi NN$ form factors.  The regularization procedure
constitutes the main model dependence in the calculation of the pion
flux.  In Sec.~\ref{sec:E866} we ask what constraints on the pion flux
models can be obtained from the SU(2) flavor asymmetry of the sea
observed in the E866 experiment.  To this end we perform a $\chi^2$
analysis for various pion distribution models, and analyze whether
any of the models can be excluded by the data.  Since the flavor
asymmetry is an inclusive observable, we consider also $\Delta$
isobar contributions in the pion--baryon dissociations, along with
the nucleon.

The HERA leading neutron data are analyzed in Sec.~\ref{sec:HERA}.
Rather than attempt to fit over the entire range of kinematics,
we restrict the analysis to the small pion momentum region where
one-pion exchange is expected to be the dominant contribution.
Since the calculations of the background processes are considerably
more model dependent, the precise delineation of the pion dominated
region is {\it a priori} unknown.  Instead of introducing additional
model dependence into the analysis, we will allow the data to
select the kinematics where pion exchange is the relevant process.
The main part of the analysis is the combined fit to the HERA and
E866 data, over a large range of $x$ and $Q^2$ values covered in
the experiments.
We discuss the impact of the E866 data on constraining models of
the pion flux, and the resulting model dependence of the extracted
pion structure function at small $x_\pi$.  Further constraints on
$F_2^\pi$ from upcoming tagged DIS experiments at Jefferson Lab at
intermediate $x_\pi$ values are discussed in Sec.~\ref{sec:TDIS},
where we illustrate how the new data may resolve some of the
differences between our fits and extrapolations of existing pion
PDFs into the low-$x_\pi$ region.
Finally, in Sec.~\ref{sec:conclusion} we summarize our findings
and suggest possible improvements in pion structure function
analyses in the future.

\section{Pion exchange models}
\label{sec:pion}

In this section we review the computation of the pion light-cone
momentum distributions in the nucleon (sometimes also referred to
as the pion splitting functions), for both $\pi N$ and $\pi \Delta$
fluctuations of the proton.
After outlining the derivation of the distributions for the case
of point particles within the framework of chiral effective theory,
we then discuss various regularization prescriptions that have been
used in the literature to regulate the ultraviolet divergences for
the more realistic case when hadron structure is taken into account.

\subsection{Pion light-cone momentum distributions}
\label{ssec:split}

\begin{figure}
\includegraphics[width=12cm]{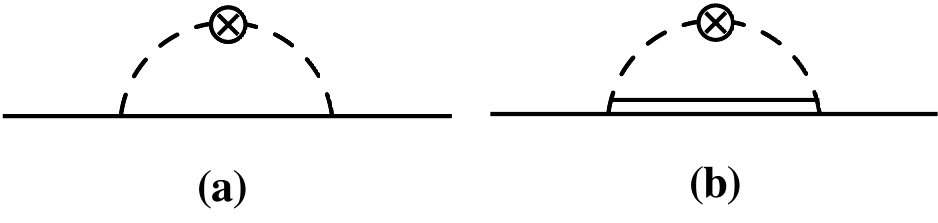}
\caption{Contributions to the pion distributions in the proton
	from the rainbow diagrams involving
	{\bf (a)} a nucleon (solid lines) and
	{\bf (b)} a $\Delta$ isobar (double solid line)
	in the intermediate state.  The external operators
	couple to the virtual pions (dashed lines).}
\label{fig:rainbow}
\end{figure}

For the fluctuation of a proton (with four momentum $p$) to a
positively charged pion (momentum $k$) and a neutron ($p-k$),
illustrated by the ``rainbow'' diagram in Fig.~\ref{fig:rainbow}~(a),
the $p \to n \pi^+$ splitting function derived from chiral
effective theory is expressed as a sum of on-shell and
$\delta$-function pieces \cite{Burkardt13, Salamu15},
\begin{equation}
f_{\pi^+ n}(y)
= 2 \Big[ f_N^{\rm (on)}(y) + f_N^{\rm (\delta)}(y) \Big],
\label{eq:fpi+n}
\end{equation}
where $y = k^+/p^+$ is the fraction of the proton’s light-cone
momentum carried by the pion, and the ``+'' component of the
four-vector is defined as $k^+ \equiv k^0 + k^z$.  The on-shell
contribution $f_N^{\rm (on)}$ corresponds to the region $y>0$
and can be written as \cite{Thomas83, DLY69}
\begin{eqnarray}
f_N^{\rm (on)}(y)
&=& \frac{g_A^2 M^2}{(4\pi f_\pi)^2}       
    \int\!dk_\perp^2\,
    \frac{y\, (k_\perp^2 + y^2 M^2)}{(1-y)^2 D_{\pi N}^2},    
\label{eq:fNon}
\end{eqnarray}
where $M$ is the nucleon mass, $g_A = 1.267$ is the axial charge,
$f_\pi = 93$~MeV is the pion decay constant, and
\begin{eqnarray}
D_{\pi N}\,
&\equiv&\, t - m_\pi^2\
 =\ -\frac{1}{1-y} \big[ k_\perp^2 + y^2 M^2 + (1-y) m_\pi^2 \big]
\label{eq:DpiN}
\end{eqnarray}
for an on-shell nucleon intermediate state, with the pion
virtuality $t \equiv k^2 = -(k_\perp^2 + y^2 M^2)/(1-y)$.
The second term in Eq.~(\ref{eq:fpi+n}), $f_N^{(\delta)}$,
arises from off-shell nucleon contributions and is proportional
to $\delta(y)$.
The significance of this term has been discussed \cite{Z1} with
respect to the model-independent nonanalytic structure of the 
vertex renormalization constant as a function of the pion mass.
One may regard this nonanalytic function of $m_\pi^2$ as the
first principles constraint on the infrared behavior of the chiral
effective theory consistent with the chiral symmetry of QCD.
In scattering processes this term contributes only at $x=0$,
and is therefore relevant only for the lowest moment of the
parton distribution.
In this work we will be analyzing data at nonzero values of $x$,
at which $f_N^{\rm (\delta)}$ will play no direct role.

Note that the factor 2 in Eq.~(\ref{eq:fpi+n}) is an isospin
factor specific to the $p \to n \pi^+$ fluctuation;
the distribution for the fluctuation $p \to p\, \pi^0$
is related to that in Eq.~(\ref{eq:fpi+n}) by
$f_{\pi^+ n}(y) = 2 f_{\pi^0 p}(y)$.
In writing the coefficient in front of the integration in
Eq.~(\ref{eq:fNon}), we have assumed the Goldberger–Treiman
relation, $g_A/f_\pi = g_{\pi NN}/M$, where
$g^2_{\pi NN}/4\pi \approx 13.7$ gives the strength of the
$\pi NN$ coupling \cite{Bugg04}.

In addition to the nucleon intermediate states, contributions from
$\Delta$ baryons in Fig.~\ref{fig:rainbow}~(b) are known to play an
important role in hadron structure.  Within the same chiral effective
theory framework, using an effective $\pi N \Delta$ interaction
\cite{Salamu15}, the $p \to \Delta^0\, \pi^+$ splitting function
can be written as a sum of three terms,
\begin{equation}
f_{\pi^+ \Delta^0}(y)
= f_\Delta^{\rm (on)}(y)
+ f_\Delta^{(\delta)}(y)
+ f_\Delta^{\rm (end\textrm{-}pt)}(y).
\label{eq:fypiD}  
\end{equation}   
The on-shell piece $f_\Delta^{\rm (on)}$, corresponding to the
$\Delta$ pole, is given for $0<y<1$ by
\begin{equation}
\!f_\Delta^{\rm (on)}(y)
= C_\Delta \int\!dk_\bot^2\,
  \frac{y\, (\overline{M}^2-m_\pi^2)}{(1-y) D_{\pi \Delta}^2}
  \Big[
    (\overline{M}^2\!-\!m_\pi^2)(\Delta^2\!-\!m_\pi^2)
  - [3(\Delta^2\!-\!m_\pi^2) + 4M M_\Delta] D_{\pi \Delta}
  \Big]
\label{eq:fDon}
\end{equation}
where
\begin{eqnarray}
D_{\pi\Delta}\,
&\equiv&\, t - m_\pi^2\
=\ -\frac{1}{1-y}
    \big[ k_\perp^2 - y(1-y) M^2 + y M_\Delta^2 + (1-y) m_\pi^2 \big]
\label{eq:DpiD}
\end{eqnarray}
for an on-shell $\Delta$ intermediate state of mass $M_\Delta$,
with $\overline{M} \equiv M_\Delta + M$ and
     $\Delta \equiv M_\Delta - M$.
The pion virtuality here is given by
$t \equiv k^2 = -(k_\perp^2 - y(1-y) M^2 + y M_\Delta^2)/(1-y)$.
The coefficient
	$C_\Delta = g_{\pi N\Delta}^2 / [(4\pi)^2 18 M_\Delta^2]$
contains the $\pi N \Delta$ coupling constant, which is given
from SU(6) symmetry by
$g_{\pi N\Delta}   
  = (3\sqrt{2}/5) g_A/f_\pi \approx 11.8$~GeV$^{-1}$ \cite{MST99}.
For the other charge channels in the $p \to \Delta \pi$
dissociation, the splitting functions are related by
$2 f_{\pi^- \Delta^{++}} = 3 f_{\pi^0 \Delta^+} = 6 f_{\pi^+ \Delta^0}$.

Note that the on-shell contribution in Eq.~(\ref{eq:fDon})
differs from the ``Sullivan'' form often used in the literature
\cite{Kumano98, Speth98, TMS00, MST99}, which is obtained by taking
the $\Delta$-pole contribution, $D_{\pi\Delta} \to M_\Delta^2$.
In particular, it has a higher power of $k_\perp$
($k_\perp^6$ compared with $k_\perp^2$ in Eq.~(\ref{eq:fDon})),
which arises from the neglect of the end-point contributions
in the Sullivan process.

The other two terms in Eq.~(\ref{eq:fypiD}), $f_\Delta^{(\delta)}$
and $f_\Delta^{\rm (end\textrm{-}pt)}$, correspond to a
$\delta$-function contribution at $y=0$ and an end-point contribution
proportional to a $\delta$-function at $y=1$, respectively.
Typically the latter term will be suppressed in the presence of a
form factor regulator, which we discuss in the next section.

Finally, for reference we also define the average multiplicities
of pions for the $\pi N$ and $\pi \Delta$ dissociations, summed
over all charge states,
\begin{subequations}
\label{eq:<n>}
\begin{eqnarray}
\label{eq:<n>piN}
\langle n \rangle_{\pi N}
&=& 3 \int_0^1 dy\, f_N^{\rm (on)}(y),	\\
\label{eq:<n>piD}
\langle n \rangle_{\pi \Delta}
&=& 6 \int_0^1 dy\, f_\Delta^{\rm (on)}(y).
\end{eqnarray}
\end{subequations}%
These will be useful for comparing the relative magnitudes of
the various models with respect to the shape of the respective
form factor regulators.

\subsection{Regularization prescriptions}
\label{ssec:FFs}

From the on-shell nucleon and $\Delta$ splitting functions in
Eqs.~(\ref{eq:fNon}) and (\ref{eq:fDon}), it is evident that
integration over contributions from large $k_\perp$ will introduce
logarithmic divergences in the point-like theory, which must be
regularized in order to obtain finite results.  Since the nucleon is
not point-like, but has a finite spatial extent of ${\cal O}$(1~fm),
this introduces an additional scale into the effective theory,
along with the chiral symmetry breaking scale \cite{FRR}.
The precise way that the finite range of the nucleon is implemented
in order to regularize the ultraviolet divergences depends on the
prescription adopted \cite{FRR, Young}, although any prescription
must correctly incorporate the infrared behavior of pion loops which
is model independent.
In practice, the model dependence amounts to a choice of form factor
$F(y,k_\perp^2)$ multiplying the integrands of Eqs.~(\ref{eq:fNon})
and (\ref{eq:fDon}) which suppresses the large-$k_\perp$
contributions.

The simplest way to regularize the integrals in the $\pi N$ and
$\pi \Delta$ splitting functions is to introduce an ultraviolet
cutoff on the $k_\perp$ integrations,
\begin{eqnarray}
F &=& \Theta(\Lambda^2-k_\perp^2)
      \hspace*{1.5cm} [k_\perp\ \textrm{cutoff}],
\label{eq:FF_kT2cut}
\end{eqnarray}
with $\Lambda$ the cutoff parameter.  Of course, a $k_\perp$
cutoff breaks Lorentz invariance, and in practice is used mainly
for illustration purposes rather than as a realistic model
for describing the momentum dependence at $k_\perp \gg 0$.
Nevertheless, as the simplest regularization prescription,
it can serve as a useful reference point with which to compare
other calculations.

Regularization prescriptions that do satisfy Lorentz invariance,
as well as chiral symmetry, include dimensional regularization and
Pauli-Villars (PV) subtraction.  For the latter, the divergence of
the amplitude is removed by subtracting from the original integrand
an amplitude with the physical pion mass replaced by a PV mass
parameter \cite{Pauli-Villars}.  Motivated by the PV regularization,
we subtract from the pion propagator $1/D_{\pi N}$ in
Eq.~(\ref{eq:fNon}) a similar term with the pion mass replaced by a
cutoff mass $\Lambda$, namely $1/D_{\pi N}^2 - 1/(t-\Lambda^2)^2$,
and similarly for the $1/D_{\pi\Delta}^2$ term in Eq.~(\ref{eq:fDon}).
This regularization method differs from the usual prescription of
introducing a form factor $F$ to each of the meson--baryon vertices,
resulting in multiplying the integrands in $f_N^{\rm (on)}$ and
$f_\Delta^{\rm (on)}$ by $|F|^2$. 
In terms of the usual prescription with form factors, our PV-motivated
regularization corresponds to introducing an effective form factor
\begin{eqnarray}
F &=& \left[1 - \frac{(t-m_\pi^2)^2}{(t-\Lambda^2)^2} \right]^{1/2}
      \hspace*{1.5cm} [\textrm{Pauli-Villars}].
\label{eq:FF_PV}
\end{eqnarray}
Note, however, that the application of the Pauli-Villars
regularization here is not unique, and other subtraction prescriptions
are possible.  For the $\pi \Delta$ case, an alternative procedure
would be to write the second term in Eq.~(\ref{eq:fDon}) as an
overall $1/D_{\pi \Delta}$, and apply the subtraction on
$1/D_{\pi \Delta}$ rather than on $1/D_{\pi \Delta}^2$.
However, since our phenomenological analysis will involve fitting
the $\Lambda$ parameter to data, it will make little difference
which we employ, and in practice we choose the latter prescription
as in Eq.~(\ref{eq:FF_PV}).

A similar regularization prescription that is often adopted in the
literature is to use a form factor that is a monopole in $t$,
\begin{eqnarray}
F &=& \left( \frac{\Lambda^2 - m_\pi^2}{\Lambda^2 - t} \right)
      \hspace*{1.5cm} [t\textrm{-dependent\ monopole}].
\label{eq:FF_mon}
\end{eqnarray}
Alternatively, a dipole form is sometimes also used, in which
the form factor is given by the square of the expression in
Eq.~(\ref{eq:FF_mon}).
A generalization of the monopole or dipole is an exponential form,
\begin{eqnarray}
F &=& \exp\big[ (t-m_\pi^2)/\Lambda^2 \big]
      \hspace*{1.5cm} [t\textrm{-dependent\ exponential}],
\label{eq:FF_exp}
\end{eqnarray}
which is an effective sum over infinitely many multipoles.
In practice, results for the dipole form factor are typically
intermediate between those for the monopole and exponential,
so using the latter two is sufficient to cover the range of
possible behaviors.

As an alternative to the $t$-dependent form factors
(\ref{eq:FF_PV})--(\ref{eq:FF_exp}), a form that naturally arises
in infinite momentum frame or light-front approaches is one in
which the form factors are functions of the invariant mass squared
of the intermediate $\pi N$ system,
$s \equiv (p+k)^2 = (k_\perp^2 + m_\pi^2)/y + (k_\perp^2 + M^2)/(1-y)$,
and similarly for the $\pi\Delta$ system with $M \to M_\Delta$.
In this case a common form is an exponential function in $s$
\cite{Zoller92, Holtmann96},
\begin{eqnarray}
F &=& \exp\big[ (M^2-s)/\Lambda^2 \big]
      \hspace*{1.5cm} [s\textrm{-dependent\ exponential}],
\label{eq:FF_LF}
\end{eqnarray}
although other $s$-dependent functional forms have also
been used in the literature \cite{MT93, MST99}.

In addition to the $s$-dependent and $t$-dependent form factors,
one may also consider $u$-dependent form factors \cite{Holtmann96}
with $u \equiv (p-k)^2 = -(k_\perp^2 - y(1-y)M^2 + y m_\pi^2)/y$
by crossing the pion virtuality to the intermediate baryon virtuality.
However, the $u$-dependent form factors are not accessible to the
on-shell contributions, $f_N^{\rm (on)}$ and $f_\Delta^{\rm (on)}$,
in which the four-momentum of the intermediate baryon is fixed by
the on-mass-shell condition.

In studies of inclusive neutron production in hadronic charge
exchange reactions, such as $h p \to n X$ ($h=\pi$ or $p$),
it was found that the exchange of Regge trajectories with pion
quantum numbers played an important role at very small values of
$y$ and finite $t$.  Within Regge theory, the pion trajectory is
incorporated through an additional effective form factor
\cite{Bishari}
\begin{eqnarray}
F &=& y^{-\alpha_\pi(t)}
      \hspace*{1.5cm} [\textrm{Bishari}],
\label{eq:FF_Bishari}
\end{eqnarray}
where $\alpha_\pi(t) \approx \alpha'_\pi\, t$, with the Regge
intercept $\alpha'_\pi \approx 1$~GeV$^{-2}$.  Once the intercept is
fixed, there are no additional parameters in this model to be varied.

A generalization of the Regge model to include additional
suppression at large $t$ was considered by Kopeliovich {\it et al.}
\cite{Kopeliovich12} in the guise of an exponential form factor
$\sim \exp(R^2\, t)$, with $R \approx 0.1$~fm.
This can be recast in a form that combines the Regge factor in
Eq.~(\ref{eq:FF_Bishari}) with the exponential form factor in
in Eq.~(\ref{eq:FF_exp}),
\begin{eqnarray}
F &=& y^{-\alpha_\pi(t)}
      \exp\big[ (t-m_\pi^2)/\Lambda^2 \big]
      \hspace*{1.5cm} [\textrm{Regge exponential}],
\label{eq:FF_Regge}
\end{eqnarray}   
with $\Lambda$ a free parameter.
We note again that in the application of each of these regularization
prescriptions in the splitting functions, it is the square of
the form factor, $|F(y,k_\perp^2)|^2$, that multiplies the
integrands in $f_N^{\rm (on)}$ and $f_\Delta^{\rm (on)}$.

\begin{figure}
\includegraphics[width=8cm]{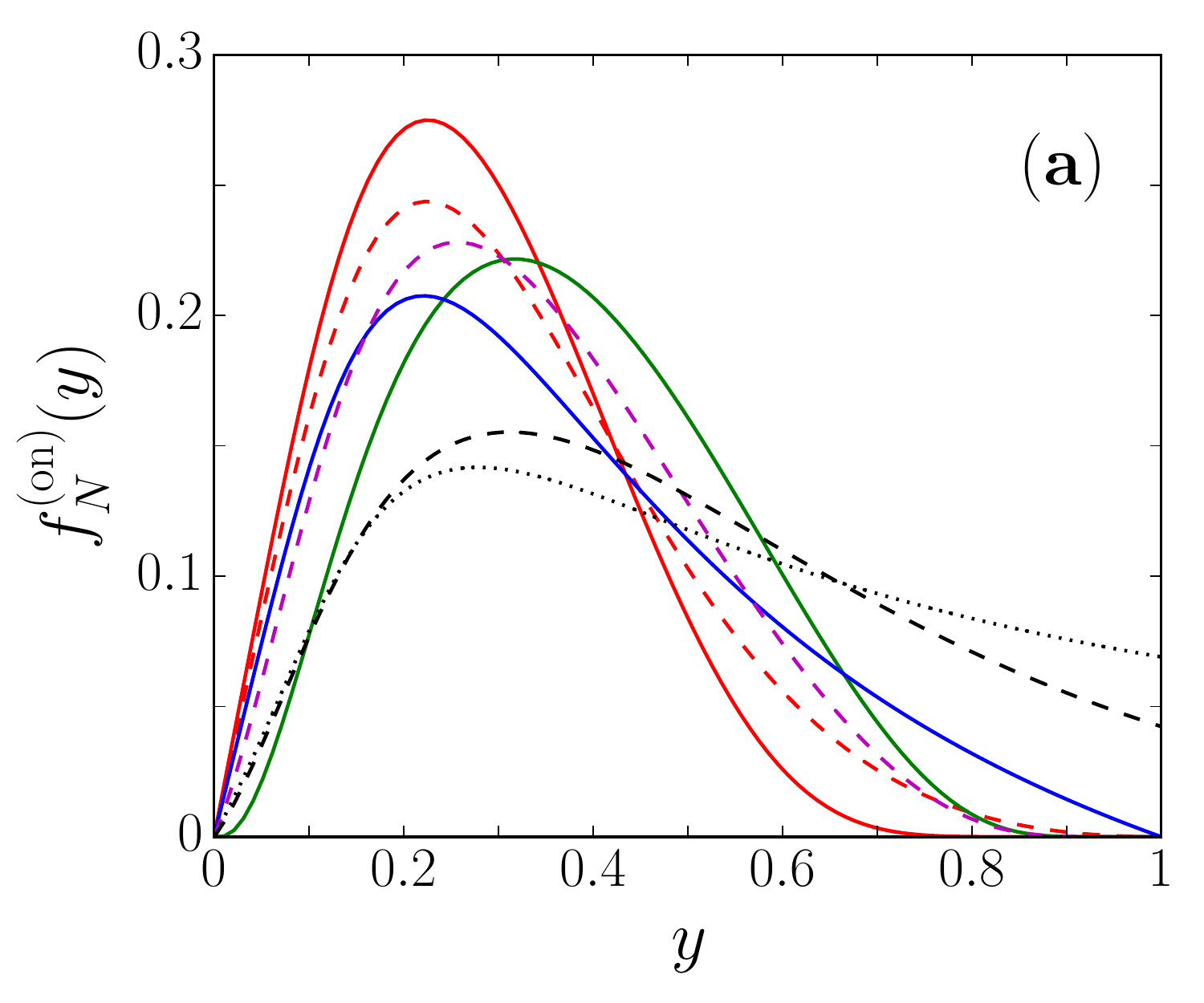}
\includegraphics[width=8cm]{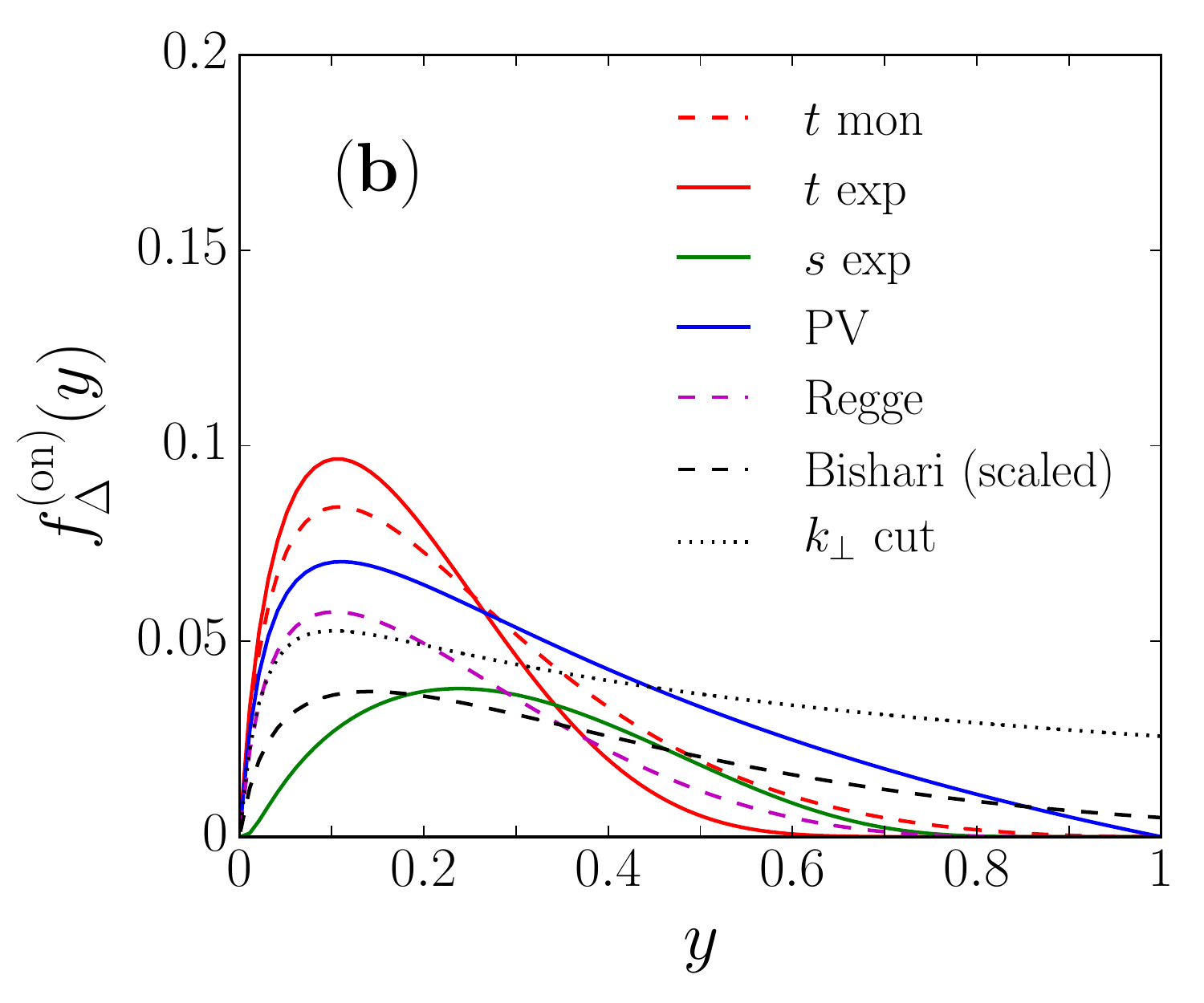}
\caption{On-shell $\pi N$ and $\pi\Delta$ splitting functions
	{\bf (a)} $f_N^{\rm (on)}(y)$ and
	{\bf (b)} $f_\Delta^{\rm (on)}(y)$ for various
	regularization prescriptions.  The $\pi N$ functions
	are normalized arbitrarily to 0.1.
	The distribution with the Bishari form factor is
	scaled down by a factor 1.9 to coincide with the same
	normalization, and the $\pi \Delta$ distributions are
	computed for the same $\Lambda$ values as the $\pi N$.}
\label{fig:fy}
\end{figure}

In Fig.~\ref{fig:fy} we illustrate the various on-shell splitting
functions $f_N^{\rm (on)}$ for the models (\ref{eq:FF_kT2cut}) --
(\ref{eq:FF_Regge}).  For reference, each of the $\pi N$ splitting
functions is normalized to 0.1 when integrated over $y$ from 0 to 1,
which for the various models corresponds to $\Lambda$ parameters of
 0.68~GeV [$t$ monopole (\ref{eq:FF_mon})],
 0.86~GeV [$t$ exponential (\ref{eq:FF_exp})],
 1.48~GeV [$s$ exponential (\ref{eq:FF_LF})],
 0.26~GeV [Pauli-Villars (\ref{eq:FF_PV})],
 1.61~GeV [Regge exponential (\ref{eq:FF_Regge})], and
 0.23~GeV [$k_\perp^2$~cutoff (\ref{eq:FF_kT2cut})].
For the Bishari model (\ref{eq:FF_Bishari}), which has no form factor
parameter beyond the Regge intercept $\alpha'_\pi$, the integrated
value of $f_N^{\rm (on)}$ is $\approx 0.19$.  To compare the shape
of this distribution with other models we normalize the splitting
function to the 0.1 value for the other functions.

The $\pi N$ splitting functions in most of the models typically have
a similar shape, increasing from $y=0$ to peak at $y \approx 0.2-0.3$.
Generally, the distributions computed with the $t$-dependent form
factors (monopole, exponential, Pauli-Villars, and Regge exponential)
are peaked at the lower $y$ values ($y \approx 0.2$), while the
additional suppression at small $y$ from the $s$-dependent form in
Eq.~(\ref{eq:FF_LF}) shifts the peak in the $s$-dependent exponential
model to larger $y$ ($y \approx 0.3$).  Without a $t$- or $s$-dependent
form factor suppression at large $k_\perp^2$, the splitting function
for the Bishari and $k_\perp$ cutoff models remains finite at $y=1$.

Similar features characterize the splitting functions for the
$\Delta$ intermediate states.  Because of the larger mass of
the $\Delta$ baryon compared to the nucleon, the peaks in the
$f_\Delta^{\rm (on)}$ functions are shifted to slightly smaller
$y$ values ($y \approx 0.1-0.2$).  The biggest difference, however,
is in the magnitude of the functions, which are $\approx 2-3$
times smaller than the nucleon $f_N^{\rm (on)}$ for the same
values of the cutoff parameters.

In the remaining part of the paper we will examine the efficacy of
the pion exchange models described in this section in fitting the
HERA leading neutron production data \cite{ZEUS_02, H1_10}, and
the compatibility of the results with the $\bar d-\bar u$ asymmetry
extracted from the E866 Drell-Yan measurement \cite{E866}.

\section{Constraints from SU(2) flavor asymmetry of the sea}
\label{sec:E866}

One of the most suggestive indirect indications of the important
role played by the pion cloud of the nucleon is the nonzero SU(2)
flavor asymmetry $\bar d - \bar u$ in the proton sea.
The first evidence for a nonzero flavor asymmetry came from the
observation by the New Muon Collaboration (NMC) of a violation
of the Gottfried sum rule \cite{NMC}, which was extracted from the
difference of proton and neutron $F_2$ structure functions over a
large range of $x$.  However, while the NMC result was the first
accurate determination of the integrated value of $\bar d-\bar u$,
extraction of its $x$ dependence required assumptions about the
shape of the valence quark PDFs which also contribute to $F_2$.
A direct determination of the $x$ dependence of $\bar d-\bar u$
was achieved through measurement of proton--proton and
proton--deuteron dimuon production cross sections in the
Drell-Yan process $pp(d) \to \mu^+ \mu^- X$ at large values of
the dimuon mass \cite{Ellis91}.

The E866 experiment at Fermilab measured the ratio
$\sigma^{pd}/\sigma^{pp}$ at high (projectile) proton momentum
fractions $x_1$ and low target momentum fraction $x_2$, where at
leading order in the strong coupling constant $\alpha_s$ it is
approximately given by \cite{E866}
\begin{eqnarray}
\frac{\sigma^{pd}}{2\sigma^{pp}}
&\approx& \frac12 \left( 1 + \frac{\bar d(x_2)}{\bar u(x_2)} \right),
\hspace*{1.5cm} [x_1 \gg x_2].
\label{eq:pd/pp}
\end{eqnarray}
The cross sections were measured for $x_2$ between 0.015 and 0.35,
at an average dimuon mass squared of $Q^2 = 54$~GeV$^2$, and the
extracted $\bar d/\bar u$ ratio was found to exceed 1.5 for
$x_2 \approx 0.1-0.2$.

In this section we examine the constraints on the models of the
pion cloud of the nucleon that can be inferred from a detailed
analysis of the $\bar d-\bar u$ asymmetry in the proton.  Within
the effective chiral framework described in Sec.~\ref{sec:pion},
the contributions to the $\bar d-\bar u$ difference from the
pion loop diagrams in Fig.~\ref{fig:rainbow} can be written as
\cite{Salamu15}
\begin{equation}
\bar d - \bar u
= \Big( f_{\pi^+ n} - \frac{2}{3} f_{\pi^-\Delta^{++}} \Big)
  \otimes \bar{q}_v^\pi,
\label{eq:conv}
\end{equation}
where $\bar{q}_v^\pi
\equiv \bar{d}^{\pi^+} - d^{\pi^+}
     = \bar{u}^{\pi^-} - u^{\pi^-}$
is the valence quark PDF in the pion, and the symbol ``$\otimes$''
denotes the convolution integral
$f \otimes q = \int_0^1 dy \int_0^1 dz\, f(y)\, q(z)\, \delta(x-yz)$.
The convolution in Eq.~(\ref{eq:conv}) follows from the crossing
symmetry properties of the splitting functions $f(-y) = f(y)$
\cite{Chen02}, and isospin symmetry relations have been assumed for
the $\pi \Delta$ distributions.  The contributions from neutral pions
cancel in the asymmetry.

\begin{figure}[t]
\includegraphics[width=16.5cm]{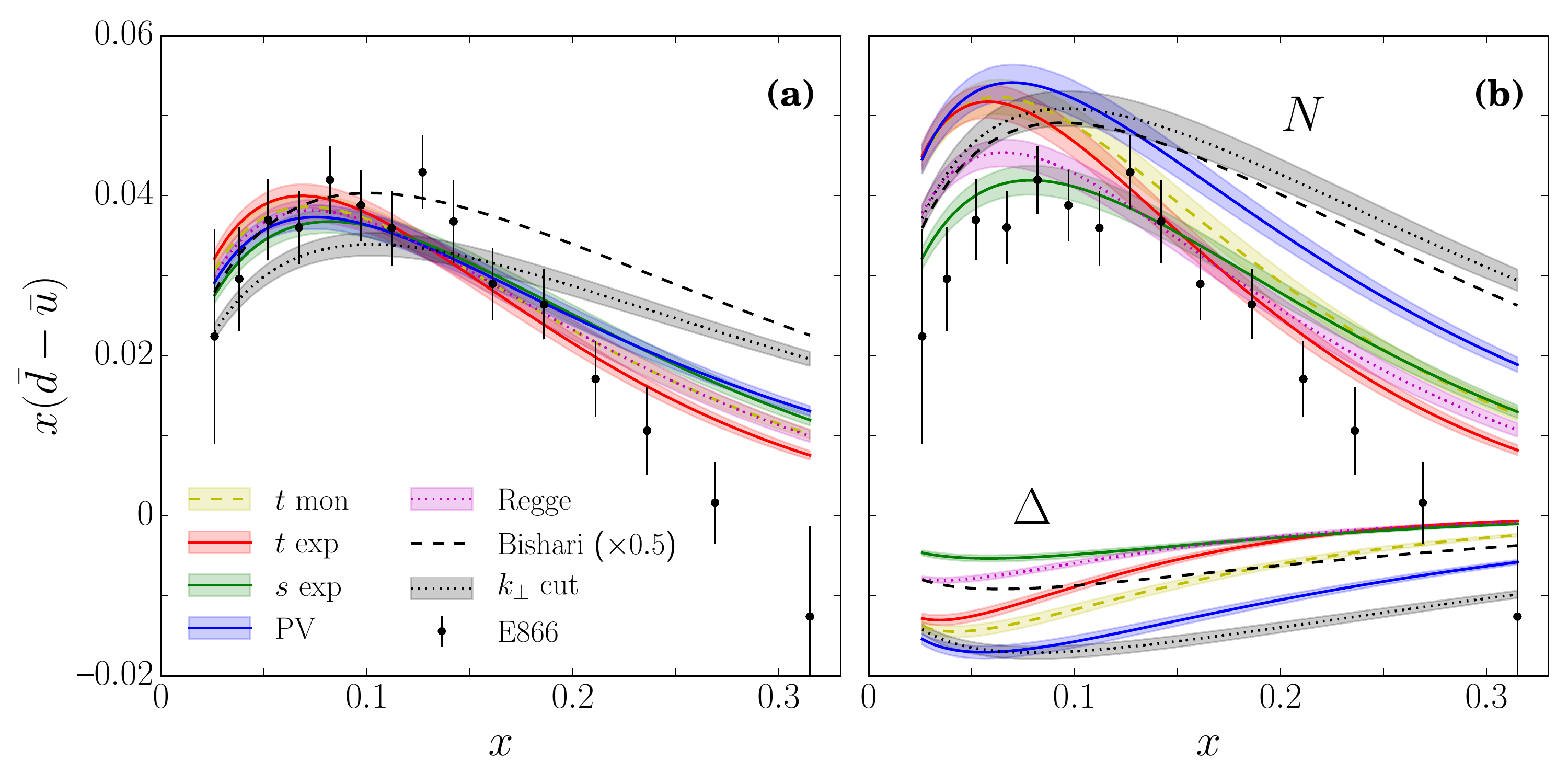}
\caption{Comparison of the flavor asymmetry $x(\bar d-\bar u)$ for
	{\bf (a)} pion model fits for various regularization
	prescriptions with the empirical asymmetry extracted
	from the E866 Drell-Yan experiment \cite{E866}, and
	{\bf (b)} the individual (positive) nucleon and
	(negative) $\Delta$ contributions to the asymmetry.
	The envelopes indicate the 68\% confidence limits.}
\label{fig:dubar_E866}
\end{figure}

\begin{table}[hbt]
\begin{center}
\caption{Best fit values for the form factor cutoffs in the
	$\pi N$ splitting function and the corresponding
	$\chi^2_{\rm dof}$ determined from the comparison with
	the $\bar d-\bar u$ asymmetry extracted from the E866
	Drell-Yan data \cite{E866}.  The associated average
	multiplicities of pions for the $\pi N$ and $\pi \Delta$
	dissociations, summed over all charge states, are also given.
	For the pion PDFs the SMRS parameterization \cite{SMRS} is
	used (the results with the ASV parameterization \cite{ASV}
	are listed in parentheses).  For the Bishari model, the
	quantities with asterisks $(^*)$ are not fitted.
	The degree of compatibility (DOC) is computed relative
	to the $t$-dependent exponential model
	(\ref{eq:FF_exp})$^\dagger$.		\\}
\begin{tabular}{l|c|c|c|c|c}\hline
\ \ \ \ model		&\ $\Lambda$ (GeV)\
			&\ $\langle n \rangle_{\pi N}$\ 
			&\ $\langle n \rangle_{\pi \Delta}$\
			&\ $\chi^2_{\rm dof}$ &\ DOC		\\ \hline
$t$ mon~	  &~0.68 (0.70)~ &~0.30 (0.32)~ &~0.18 (0.23)~
				 &~1.4 (1.2)~   &~~60\% (55\%)	\\
$t$ exp~	  &~0.85 (0.88)~ &~0.29 (0.31)~ &~0.16 (0.17)~
				 &~1.2 (1.1)~   &~100\% (100\%)$^\dagger$\\
$s$ exp~	  &~1.33 (1.36)~ &~0.23 (0.24)~ &~0.06 (0.07)~
				 &~1.8 (1.3)~   &~~24\% (19\%)	\\
Pauli-Villars~	  &~0.27 (0.27)~ &~0.31 (0.33)~ &~0.21 (0.23)~
				 &~1.9 (1.5)~   &~~30\% (23\%)	\\
Regge exp~	  &~1.32 (1.41)~ &~0.25 (0.27)~ &~0.10 (0.11)~
				 &~1.4 (1.1)~   &~~54\% (47\%)	\\
$k_\perp$ cutoff  &~0.23 (0.24)~ &~0.29 (0.31)~ &~0.22 (0.23)~
				 &~3.7 (3.2)~   &~~~1\% (0.5\%)	\\
Bishari		  &  ---         &~0.56$^*$~    & 0.23$^*$~
				 &  76 (67)     & ---		\\ \hline
\end{tabular}
\label{tab:E866}
\end{center}
\end{table}

Performing a $\chi^2$ fit to the E866 data, the results for the
various regularization prescriptions are compared in
Fig.~\ref{fig:dubar_E866}~(a), with the best fit cutoff parameters
and $\chi^2_{\rm dof}$ values summarized in Table~\ref{tab:E866}.
For reference, we also list in Table~\ref{tab:E866} the values of
the average multiplicities of pions for the $\pi N$ and
$\pi \Delta$ dissociations from Eqs.~(\ref{eq:<n>}).
The uncertainty bands around the central values for each of the
models have been computed using standard Hessian error analysis,
as described in Appendix~\ref{sec:Hessian}.
For the valence antiquark distribution in the pion we use the
SMRS parametrization \cite{SMRS} of the world's data from $\pi N$
Drell-Yan and prompt photon production, evaluated at the E866
average $Q^2$ of 54~GeV$^2$.
In all the fits the same cutoff parameters have been taken for
the $\pi N$ and $\pi \Delta$ splitting functions, and the
individual (positive) $N$ and (negative) $\Delta$ contributions
are shown in Fig.~\ref{fig:dubar_E866}~(b).
Since the $\pi N$ and $\pi \Delta$ dissociations contribute to the
asymmetry with opposite signs, allowing these to vary independently
leads to very large correlations, as different combinations of
$\pi N$ and $\pi\Delta$ cutoffs give essentially the same
$\bar d-\bar u$ asymmetry.
On the other hand, because the shapes of the $f_N^{\rm (on)}$ and
$f_\Delta^{\rm (on)}$ functions are different [see Fig.~\ref{fig:fy}],
more precise data on $\bar d-\bar u$ as a function of $x$ could in
future allow the $N$ and $\Delta$ contributions to be constrained
independently.

In the present fits, the values of $\langle n \rangle_{\pi N}$
range from 0.23 for the $s$-dependent form factor
[Eq.~(\ref{eq:FF_LF})] to 0.31 for the Pauli-Villars regularization
[Eq.~(\ref{eq:FF_PV})].  For the same values of the $\pi\Delta$
and $\pi N$ cutoffs, the corresponding $\pi\Delta$ multiplicities
$\langle n \rangle_{\pi \Delta}$ range from 0.06 to 0.21.
The fits with the lowest $\chi^2_{\rm dof}$ values are obtained
with the $t$-dependent exponential regulator [Eq.~(\ref{eq:FF_exp})],
although, with the exception of the Bishari [Eq.~(\ref{eq:FF_Bishari})]
and $k_\perp$ cutoff [Eq.~(\ref{eq:FF_kT2cut})] regulators, each of
the models gives a reasonable overall description of the E866 data.

For the Bishari model, in which there is no form factor parameter
other than the Regge intercept $\alpha'_\pi$, the result in
Fig.~\ref{fig:dubar_E866} represents a prediction rather than a fit.
The predicted asymmetry is therefore about two times larger than
the $\bar d-\bar u$ data (the calculation is scaled down in
Fig.~\ref{fig:dubar_E866} by a factor 2 for clarity).
Since the Bishari model was constructed to describe neutron
production in hadronic reactions at low $|t|$, it is not surprising
that when applied to a $t$-integrated quantity such as $\bar d-\bar u$
it would not give a good fit ($\chi^2_{\rm dof} \approx 76$).
Similarly, the $\chi^2$ values for the sharp $k_\perp$ cutoff
regularization are significantly larger than those for all other
fitted results ($\chi^2_{\rm dof} > 3$).  However, since this model
has been used recently in the literature to study the chiral
properties of pion loops \cite{Salamu15, Z1}, it is useful to
include it here for reference.

Note that the biggest contributions to the $\chi^2$ arise from the
high-$x$ data points, which have a steeper fall-off than can be
accommodated in any of the models.
(The new SeaQuest experiment at Fermilab \cite{SeaQuest} will
in the near future check the high-$x$ behavior by measuring the
$\bar d/\bar u$ ratio up to $x \approx 0.45$.)
If one were to fit only the points below $x \approx 0.2$, 
all of the models (apart from Bishari and $k_\perp$ cutoff) would
be essentially indistinguishable, with $\chi^2_{\rm dof} < 1$
for each.

\begin{figure}[t]
\includegraphics[width=12cm]{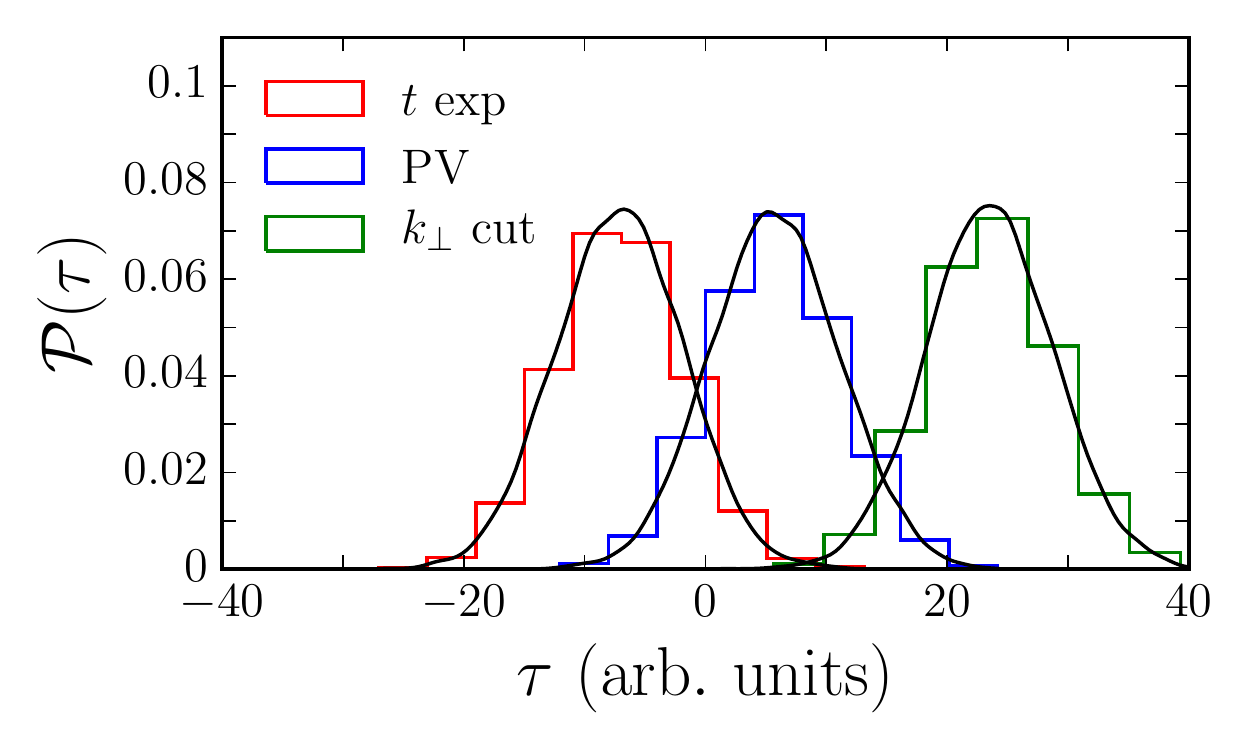}
\caption{Probability distributions ${\cal P}(\tau)$ for the
	t-statistic $\tau$ in Eq.~(\ref{eq:t-statistic})
	for the $t$-dependent exponential (best fit, red),
	PV (blue) and $k_\perp$ cutoff (worst fit, green) models.
	The units along the abscissa are arbitrary.
	The overlap between any two distributions defines
	the degree of compatibility between the models.}
\label{fig:t-dist}
\end{figure}

On the other hand, it is evident from Fig.~\ref{fig:dubar_E866}~(a)
that in some cases, in both the small-$x$ and large-$x$ regions,
the error bands on the model curves do not overlap.
To quantify the extent to which the models are compatible
amongst themselves, we employ a hypothesis test using standard
t-statistics, as described in Appendix~\ref{sec:t-statistics}.
For the null hypothesis we take the $t$-dependent exponential model
(best fit to the E866 data) and the $k_\perp$ cutoff model as
the alternative hypothesis (worst fit).
The t-distributions of the pseudodata generated from several
of the models ($t$-dependent exponential, PV and $k_\perp$ cutoff)
are shown in Fig.~\ref{fig:t-dist} for illustration.
The degree of compatibility (DOC) of each model with respect
to the best fit model ($t$-dependent exponential) is shown in
Table~\ref{tab:E866}.  From the definition, the DOC for the 
$t$-dependent exponential model is 100\%.  The DOC values for the
$t$-dependent monopole and Regge exponential models are $> 50\%$,
while, not surprisingly, for the $k_\perp$ cutoff (worst fit)
model the DOC is 1\%.

In the preceeding analysis we have examined the sensitivity of the
calculated $\bar d-\bar u$ asymmetry to the choice of model for the
hadronic pion--nucleon form factor in the pion splitting functions
$f_N^{\rm (on)}$ and $f_\Delta^{\rm (on)}$.
While the pion PDFs at small $x$ values have never been directly
measured, in the valence quark region the $\pi N$ Drell-Yan data
\cite{NA3, NA10, E615} provide strong constraints on the $x$
dependence of $\bar q_v^\pi$ for $x \gtrsim 0.1$.
Interestingly, the distributions at $x \to 1$ were observed
\cite{E615} to be more consistent with a $\sim (1-x)$ behavior
\cite{WMdual, Shigetani93, Szczepaniak94} than with the
$\sim (1-x)^2$ expectation from perturbative QCD \cite{FJ79}
or model calculations using the Dyson-Schwinger equations
(see Ref.~\cite{Holt10}).  The large-$x$ behavior in the SMRS
parametrization \cite{SMRS} was consistent with the $\sim (1-x)$
form indicated by the data.

\begin{figure}[t]
\includegraphics[width=12cm]{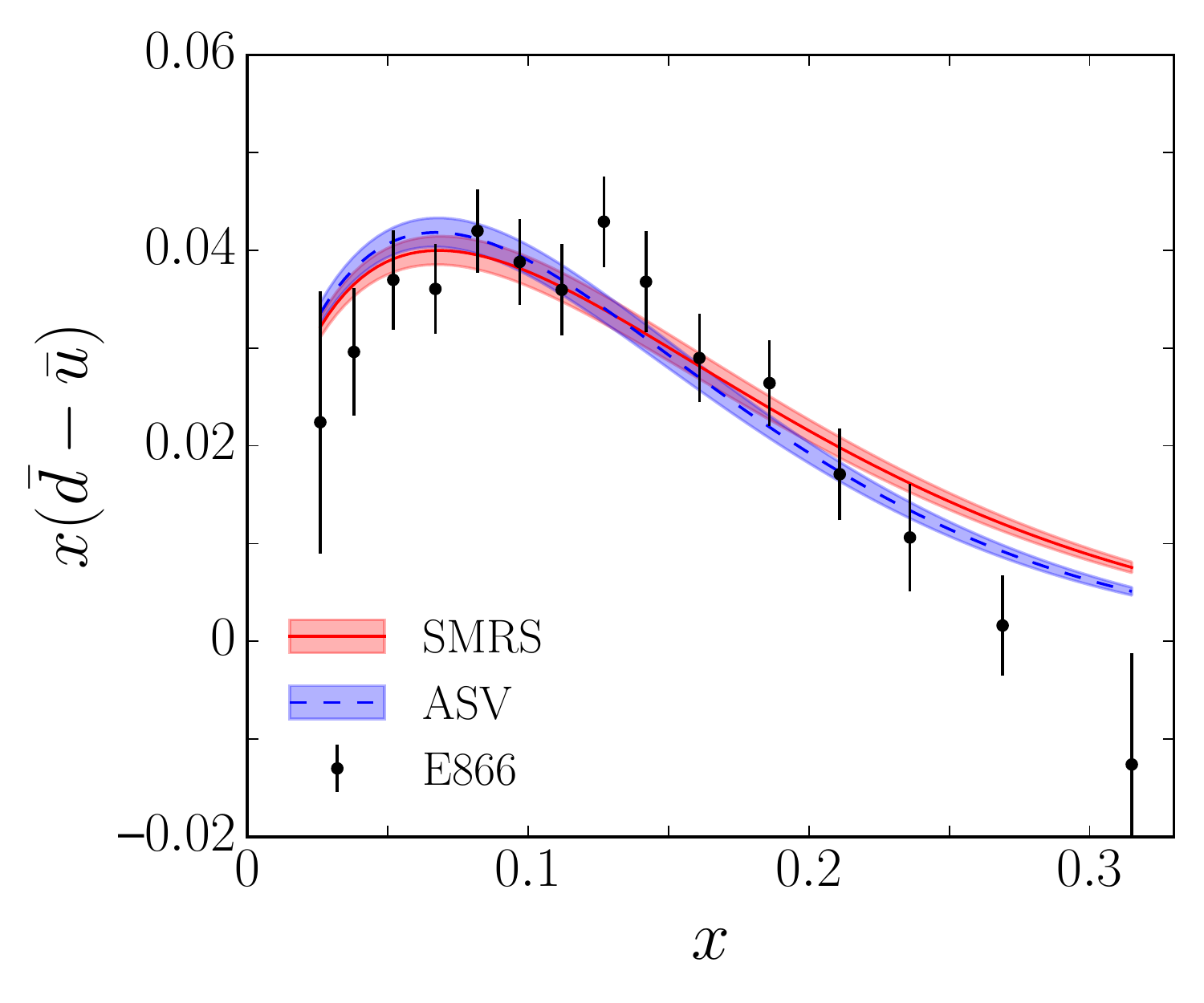}
\caption{Comparison of the pion model fits to the $\bar d-\bar u$
	data from the E866 experiment \cite{E866} with the
	$t$-dependent exponential form factor (\ref{eq:FF_exp})
	for the valence pion PDFs from the SMRS \cite{SMRS}
	and ASV	\cite{ASV} parametrizations.}
\label{fig:dubar_E866_qpi}
\end{figure}

Later, an analysis including next-to-leading order (NLO)
corrections \cite{Wijesooriya05} found that the higher order
effects soften the distributions, leading to a behavior that
was intermediate between $(1-x)$ and $(1-x)^2$.
More recently, Aicher {\it et al.} (ASV) \cite{ASV} found that
inclusion of threshold resummation at next-to-leading log (NLL)
accuracy produces valence distributions that behave approximately
as $(1-x)^2$ at a low energy scale $Q_0 = 0.63$~GeV.

In order to assess the possible impact of the different $x \to 1$
behaviors of the valence pion PDF on the $\bar d-\bar u$ asymmetry,
we repeat our analysis using the ASV parametrization \cite{ASV},
evolved from the low energy scale $Q_0$ to $Q^2=54$~GeV$^2$.
The best fit results for the various models are listed in
Table~\ref{tab:E866}, and compared in Fig.~\ref{fig:dubar_E866_qpi}
for the $t$-dependent exponential form factor (\ref{eq:FF_exp})
with the result using the SMRS parametrization.
As expected, the result with the ASV distribution leads to a
softer asymmetry, with slightly better agreement at large $x$
but marginally worse at $x \lesssim 0.1$.
The overall $\chi^2_{\rm dof}$ values are slightly better
for the ASV fit, mostly because the softer distribution allows
a smaller asymmetry at $x \gtrsim 0.2$, as preferred by the
E866 data, although the differences are not significant.
The new results for the flavor asymmetry from the SeaQuest
experiment \cite{SeaQuest} at large $x$ may provide further
insights into these comparisons.

\newpage
\section{Leading neutron production at HERA}
\label{sec:HERA}

Recently, the ZEUS \cite{ZEUS_02} and H1 \cite{H1_10} Collaborations
at HERA measured the production of neutrons in the semi-inlusive
process $e p \to e n X$, with the leading neutron carrying a large
fraction of the proton beam's momentum.
Within the one-pion exchange framework, the data were analyzed with
the aim of extracting the pion structure function $F_2^\pi$ at
small values of the pion's momentum fraction $x_\pi$
($x_\pi \gtrsim 10^{-4}$).  The previous $\pi N$ Drell-Yan
measurements \cite{E615} of the pion PDFs extended down to
$x_\pi \approx 0.2$, so that the sea quark PDFs in the pion
were essentially unconstrained.

Of course, since the leading neutron cross section in pion-exchange
models is a product of the pion structure function and the pion flux,
the extracted $F_2^\pi$ must depend to some extent on the input used
for the latter \cite{D'Alesio00, Kopeliovich12}.
The ZEUS analysis \cite{ZEUS_02} indeed suggested strong dependence
(up to a factor $\sim 2$) of $F_2^\pi$ on the model of the pion flux
adopted.  Motivated by the Regge model descriptions of inclusive
neutron spectra in $h p \to n X$ reactions, the ZEUS study
\cite{ZEUS_02} used the Bishari model (\ref{eq:FF_Bishari})
as a baseline for the analysis of the $e p$ leading neutron data,
and contrasted this with a simple additive quark model based on
constituent quark counting.
In the more recent analysis by Kopeliovich {\it et al.}
\cite{Kopeliovich12}, the Regge theory-inspired exponential
vertex function in Eq.~(\ref{eq:FF_Regge}) was employed,
while the earlier work of D'Alesio \& Pirner \cite{D'Alesio00}
considered the $t$-dependent exponential (\ref{eq:FF_exp}) and
$s$-dependent (\ref{eq:FF_LF}) forms, as well as a nontraditional
form factor extracted from Skyrme models of the $NN$ force
\cite{Holzwarth97, Fries98}.

In the present analysis we build upon these earlier studies by
systematically investigating the dependence of the fitted pion
structure function on the models of the pion splitting function,
and whether the dependence can be reduced by imposing additional
constraints from the E866 data.  The combined analysis may provide
insights into the applicability of specific functional forms,
some of which may be more attuned to describing the disparate
reactions than others.
It is also known from previous studies \cite{Nikolaev97, D'Alesio00,
H1_01, Kopeliovich12} that rescattering and absorptive effects can
play an important role in inclusive hadron production reactions.
The effects of absorption are generally found to be stronger in
$pp$ scattering than in photon-induced reactions, and decrease in
magnitude with increasing photon virtualities.  The absorptive
corrections are smaller in DIS kinematics, contributing $\sim 10\%$
at low values of $y$.
Furthermore, background contributions from other processes, such as
the exchange of heavier mesons, become increasingly more important
at larger $y$ ($y \gg 0.1$) \cite{MT93, Holtmann96, Kopeliovich12}.

\subsection{Leading neutron cross sections}
\label{ssec:sigLN}

At tree level the differential cross section for the production of
leading neutrons (LN) in semi-inclusive $ep$ scattering is given by
\begin{eqnarray}
\frac{d^3\sigma^{\rm LN}}{dx\, dQ^2\, dy}
&=& {\cal K}\, F_2^{\rm LN(3)}(x,Q^2,y),
\label{eq:d3sig}
\end{eqnarray}
where the kinematic factor
\begin{eqnarray}
{\cal K} &=& \frac{4\pi\alpha^2}{x Q^4}
	     \left( 1 - y_e + \frac{y_e^2}{2} \right),
\label{eq:K}
\end{eqnarray}
and $y_e = q \cdot p / l \cdot p \approx Q^2/xs$ is the lepton
inelasticity.  Here $l$ and $q$ are the incident lepton and
virtual photon momenta, respectively, $\alpha$ is the electromagnetic
fine structure constant, and $\sqrt{s}\, \sim\, 300$~GeV is the total
$ep$ HERA center of mass energy.
In writing Eq.~(\ref{eq:d3sig}) we have also neglected possible
contributions from rescattering and absorption.
Because in the HERA experiments the scattering angle of the
forward neutron is not measured, its transverse momentum
$p^n_\perp \approx x_L\, E_p\, \theta_n$ must be integrated over,
where $E_p$ is the energy of the incident proton beam and
$x_L = 1-y$ is the light-cone momentum fraction of the proton
carried by the neutron.
The tagged neutron structure function $F_2^{\rm LN(3)}$ is then
given by the $p^n_\perp$-integrated differential structure
function
\begin{eqnarray}
F_2^{\rm LN(3)}(x,Q^2,y)
&=& \int dp^n_\perp\, F_2^{\rm LN(4)}(x,Q^2,y,p^n_\perp).
\label{eq:F2LN3_def}
\end{eqnarray}
In the pion-exchange model the magnitude of the transverse momentum
of the leading neutron is equivalent to that of the exchanged pion,
$p^n_\perp = k_\perp$, and the fully differential structure function
$F_2^{\rm LN(4)}$ can be written in the factorized form
\begin{eqnarray}
F_2^{\rm LN(4)}(x,Q^2,y,k_\perp)
&=& 2 f_N^{\rm (on)}(y, k_\perp)\, F_2^\pi(x_\pi, Q^2),
\label{eq:F2LN4}
\end{eqnarray}
where $x_\pi = x/y$ is the fraction of momentum of the pion carried
by the interacting parton, and the pion structure function has been
assumed to be independent of $k_\perp$.
The latter assumption allows the $k_\perp$-unintegrated pion flux to
be related to the on-shell ($y > 0$) part of the splitting function
in Eq.~(\ref{eq:fNon}),
\begin{eqnarray}
f_N^{\rm (on)}(y)
&=& \int dk_\perp\, f_N^{\rm (on)}(y, k_\perp),
\end{eqnarray}
so that the tagged neutron structure function $F_2^{\rm LN(3)}$
can be written
\begin{eqnarray}
F_2^{\rm LN(3)}(x,Q^2,y)
&=& 2 f_N^{\rm (on)}(y)\, F_2^\pi(x_\pi, Q^2).
\label{eq:F2LN3}
\end{eqnarray}

The H1 experiment \cite{H1_10} measured $F_2^{\rm LN(3)}$
over a large range of kinematics covering
\mbox{$1.5 \times 10^{-4} \leqslant x \leqslant 3 \times 10^{-2}$}
and $6 \leqslant Q^2 \leqslant 100$~GeV$^2$ for average $y$
values between 0.05 and 0.68, and $p^n_\perp < 0.2$~GeV.
A similarly extensive range of kinematics was covered
by the ZEUS data \cite{ZEUS_02}, for
\mbox{$1.1 \times 10^{-4} \leqslant x \leqslant 3.2 \times 10^{-2}$}
from photoproduction up to $Q^2 \sim 10^3$~GeV$^2$, with
$0 < y < 0.8$ and neutron scattering angle $\theta_n < 0.8$~mrad.
The latter corresponds to a transverse momentum acceptance of
$p^n_\perp < 0.656~(1-y)$~GeV.  To reduce many of the correlated
systematic errors, the ZEUS experiment measured the ratio $r$
of leading neutron to inclusive cross sections in bins of width
$\Delta y$,
\begin{eqnarray}
r(x,Q^2,y)
&=& \frac{d^3\sigma^{\rm LN} / dx\, dQ^2\, dy}
	 {d^2\sigma^{\rm inc} / dx\, dQ^2}\
    \Delta y,
\label{eq:r_def}
\end{eqnarray}
where the corresponding inclusive cross section,
\begin{eqnarray}
\frac{d^2\sigma^{\rm inc}}{dx\, dQ^2}
&=& {\cal K}\, F_2^p(x,Q^2),
\end{eqnarray}
is expressed in terms of the proton structure function $F_2^p$.
In the pion exchange model $r$ is then proportional to the ratio
of the pion to proton structure functions, evaluated at $x_\pi$
and $x$, respectively,
\begin{eqnarray}
r(x,Q^2,y)
&=& 2 f_N^{\rm (on)}(y)\,
    \frac{F_2^\pi(x_\pi, Q^2)}{F_2^p(x,Q^2)}\,
    \Delta y.
\label{eq:r}
\end{eqnarray}
Multiplying the $r$ ratios by a fit to the inclusive $F_2^p$ data,
the ZEUS Collaboration was also able to reconstruct $F_2^{\rm LN(3)}$
values for various bins of $x$, $Q^2$ and $y$.

\subsection{Optimizing sensitivity to one-pion exchange}
\label{ssec:ycut}

While some dedicated analyses \cite{Kopeliovich12} have attempted
to describe the HERA leading neutron spectra at all kinematics,
our aim here will instead be to maximize the sensitivity to the
basic one-pion exchange contribution, which has the most direct
connection to the chiral effective theory.
This can be achieved by restricting the analysis to regions where
one-pion exchange is expected to be the dominant process,
and contributions from other backgrounds are minimal.
In practice, since the calculation of the backgrounds is significantly
more model dependent, the exact choice of kinematics may be somewhat
subjective.  To determine in a more objective way the region of
kinematics where the one-pion exchange is applicable, we perform
a $\chi^2$ analysis of the data as a function of the maximum value
of $y$ up to which the data are fitted.
Although this reduces the total number of data points in the fit,
the analysis of the more restrictive kinematic range should allow
for a cleaner interpretation and extraction of the pion exchange
parameters.

In performing the $\chi^2$ fits to the ZEUS \cite{ZEUS_02} and
H1 \cite{H1_10} data, for each of the models of the pion flux
discussed in Sec.~\ref{ssec:FFs} we vary the cutoff parameter
$\Lambda$ in the form factor (with the exception of the Bishari
model, which does not have a cutoff), as well as the pion
structure function.
For the pion structure function parametrization at the input
scale $Q_0^2$ we use form
\begin{eqnarray}
F_2^\pi(x_\pi,Q_0^2)
&=& N\, x_\pi^a\, (1-x_\pi)^b,
\label{eq:F2pi}
\end{eqnarray}
which should be sufficiently flexible for describing the
small-$x_\pi$ region.  Since the HERA data are insensitive
to the large-$x_\pi$ behavior of $F_2^\pi$, we fix the
parameter $b = 1$ \cite{SMRS};
the exact value of $b$ does not affect the determination
of the more relevant small-$x_\pi$ parameters, namely,
the exponent $a$ and the normalization $N$.
To allow for the $Q^2$ dependence of $a$ we use the
simple {\it ansatz} \cite{JFO84}
\begin{eqnarray}
a &=& a_0 + a_1\, \eta,
\end{eqnarray}
where the $Q^2$ dependence is parametrized through the variable
\cite{JFO84}
\begin{eqnarray}
\eta &=& \log\left( \frac{\log Q^2/\Lambda_{\rm QCD}^2}
			 {\log Q_0^2/\Lambda_{\rm QCD}^2}
	     \right),
\label{eq:eta}
\end{eqnarray}
with $Q_0^2=1$~GeV$^2$ and $\Lambda_{\rm QCD}=0.4$~GeV.
The $\eta$ dependence of $a$ effectively mocks up the
$Q^2$ evolution of the sea quark distributions in the pion.
The fits then involve a total of four parameters for each
model of the pion flux.
In principle one could also decompose $F_2^\pi$ in a partonic
representation and fit the individual valence and sea quark
PDFs in the pion, in the context of a global QCD fit
\cite{SMRS, GRV, GRS}.  Although this be a worthwhile
future pursuit, it is somewhat outside of the scope of
the present analysis.

\begin{figure}[t]
\includegraphics[width=22cm]{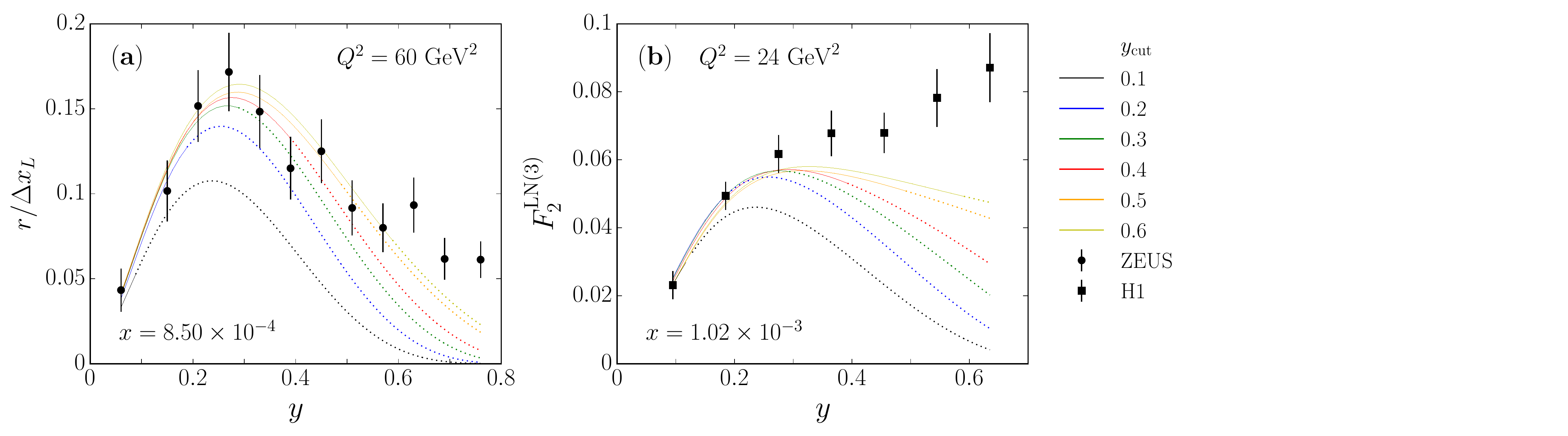}
\caption{Examples of $y$-dependent spectra of leading neutrons from
	{\bf (a)} ZEUS $r/\Delta y$ data at $x=8.5 \times 10^{-4}$
	and $Q^2=60$~GeV$^2$, and
	{\bf (b)} H1 $F_2^{\rm LN(3)}$ data at
	$x=1.02 \times 10^{-3}$	and $Q^2=24$~GeV$^2$.
	The curves represent simultaneous fits to ZEUS and H1 data
	at all available $x$ and $Q^2$ values, for fixed maximum
	values of $y$ from $y_{\rm cut} = 0.1$ to 0.6, using the
	$t$-dependent exponential form factor model (\ref{eq:FF_exp}).
	The dotted curves are extrapolations of the respective
	fits into the unconstrained regions above $y_{\rm cut}$.}
\label{fig:yspectra}
\end{figure}

To illustrate the effects on the fits to the HERA data of the
$y$ cut, we show in Fig.~\ref{fig:yspectra} the ZEUS cross section
ratio $r/\Delta y$ and the H1 $F_2^{\rm LN(3)}$ structure function
in two representative bins at fixed $x$ and $Q^2$ values.
For the ZEUS ratio $r$, we divide the calculated model
$F_2^{\rm LN(3)}$ by the proton structure function $F_2^p$
computed from the NLO PDFs parametrized in the HERAPDF1.5 set
\cite{HERAPDF1.5}.
Since the model with the $t$-dependent exponential form factor gave
the best results for the E866 data comparison in Sec.~\ref{sec:E866},
we use this model here to illustrate the $y_{\rm cut}$ dependence.
Other models give qualitatively similar results.
While the low-$y$ data can be described within the model reasonably
well, fitting the cross sections at higher $y$ values becomes
increasingly difficult.  This is not surprising, since contributions
from processes other than one-pion exchange are known to become
progressively more important with increasing $y$.
Similar behavior is seen for the $y$-dependent spectra in
other $x$ and $Q^2$ bins.
Note also that the ratio $r$ for the ZEUS data decreases beyond
$y \approx 0.3$, while $F_2^{\rm LN(3)}$ from H1 keeps increasing
with $y$ (the relative factor of $F_2^p$ between them is independent
of $y$).  The different behavior of these spectra reflects the
different detector acceptances in the two experiments with
relation to the neutron transverse momentum $p_\perp^n$.
While H1 applied a $y$-independent cut on $p_\perp^n$, the ZEUS
cut proportional to $1-y$ suppresses contributions from larger
$y$ values.

Of course, in general we would like to maximize the $y$ coverage
included in the analysis in order to increase the statistics of
the fit.  For the smallest $y$ cut, for instance, $y_{\rm cut}=0.1$,
there is a total of 54 data points (25 from ZEUS and 29 from H1),
while for $y_{\rm cut}=0.2$ the number of points doubles to 108.
For $y_{\rm cut}=0.3$, the number of points increases to 187
(100 from ZEUS and 87 from H1), and at $y_{\rm cut}=0.4$ it
reaches 266.
Furthermore, increasing the value of $y_{\rm cut}$ allows one
to maximize the range of $x_\pi$ over which the pion structure
function is constrained.
At fixed $x$, a smaller value of $y_{\rm cut}$ will restrict
the sensitivity of the fit to small $x_\pi$ values.
For example, for the ZEUS data the lowest $x$ bin extends
to $x=1.1 \times 10^{-4}$, so that a $y_{\rm cut}$ of
$\approx 0.1$ or 0.3 will allow one to reach down to
$x_\pi^{\rm min} \approx 1 \times 10^{-3}$ or $4 \times 10^{-4}$,
respectively.
In the case of the H1 data, for which the smallest $x$ value
is $2.24 \times 10^{-4}$, sensitivity to the pion structure
function can be extended down to
$x_\pi^{\rm min} \approx 2 \times 10^{-3}$ and $7 \times 10^{-4}$
for the same respective $y_{\rm cut}$ values.

\begin{figure}[t]
\includegraphics[width=16cm]{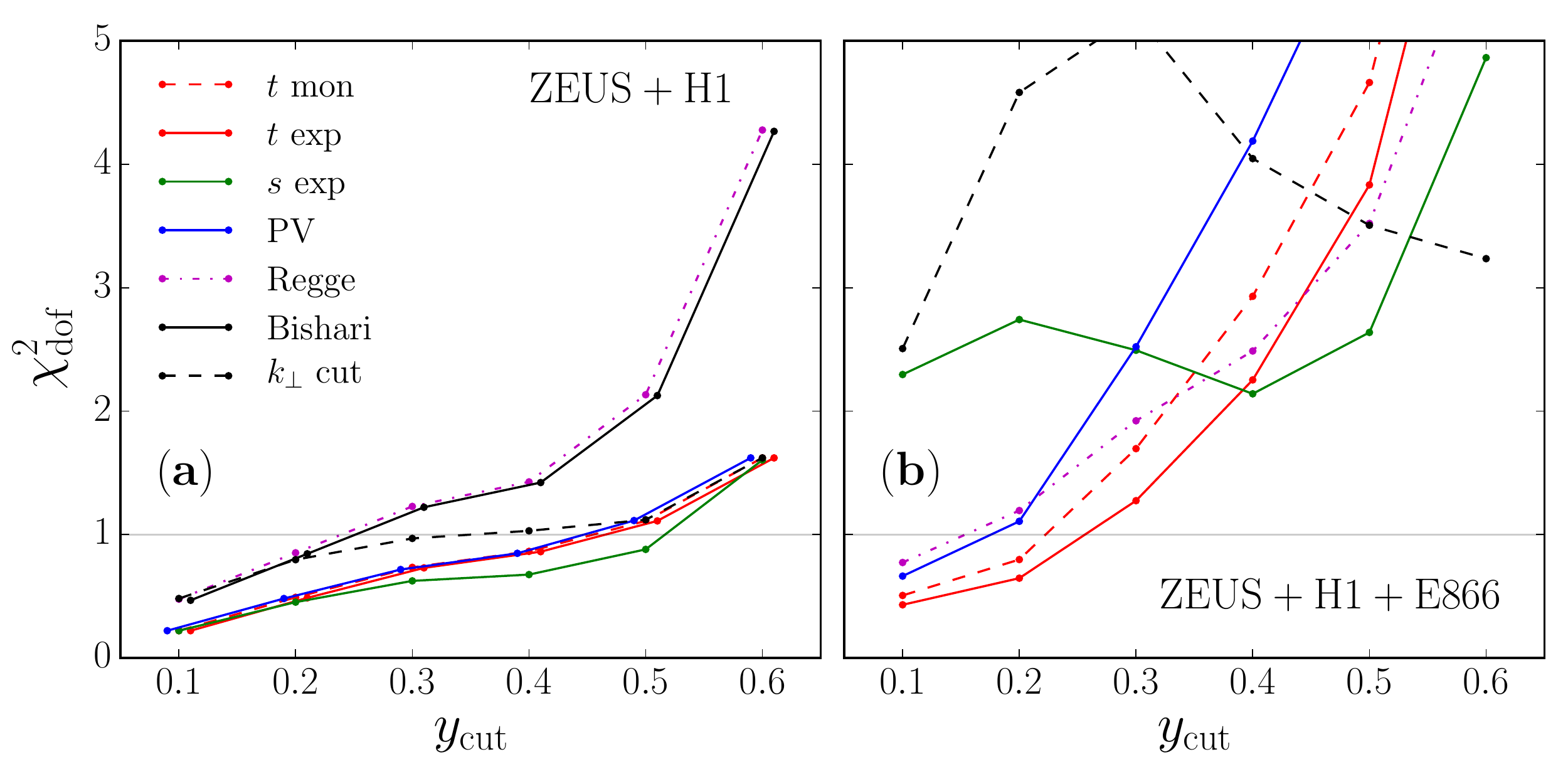}
\caption{Variation of $\chi^2_{\rm dof}$ in various models with
	the maximum value $y_{\rm cut}$ used in the fit to the
	HERA leading neutron data, for
	{\bf (a)} ZEUS \cite{ZEUS_02} and H1 \cite{H1_10} data
	only, and
	{\bf (b)} the combined ZEUS, H1 and E866 \cite{E866} fit.
	The Bishari model in the latter is off the vertical scale.}
\label{fig:chi2}
\end{figure}

To determine the sensitivity of the fit to different $y$ cuts,
we compute the $\chi^2$ values for each of the models by fitting
the ZEUS and H1 data over the respective ranges from $y=0$ to
$y_{\rm cut}$.
The $\chi^2_{\rm dof}$ profiles in Fig.~\ref{fig:chi2}~(a) for
the HERA fit indicate generally good fits for all models, with
$\chi^2_{\rm dof} \lesssim 1$ up to $y_{\rm cut} \approx 0.3$.
In fact, all the models other than the Bishari (\ref{eq:FF_Bishari})
and Regge exponential (\ref{eq:FF_Regge}) model give good
$\chi^2_{\rm dof}$ values up to $y_{\rm cut} \approx 0.5$,
above which the fits rapidly deteriorate.

A closer inspection of the fitted parameters, however, reveals
rather large correlations between the $\Lambda$ values and the pion
structure function parameters, especially for low $y_{\rm cut}$.
For example, there is a 36\% correlation between $\Lambda$ and
the normalization $N$ for $y_{\rm cut} = 0.3$, and an even larger,
51\% correlation for $y_{\rm cut} = 0.2$.
This suggests that while reasonable fits to the leading neutron
cross sections can be obtained within most of the pion exchange
models, meaningful extraction of pion structure function parameters
from the HERA data alone is problematic.  To determine the pion
parameters unambiguously requires additional constraints beyond
the leading neutron cross sections.
An obvious candidate for an independent constraint is the
$\bar d-\bar u$ asymmetry extracted from the E866 Drell-Yan data,
discussed in Sec.~\ref{sec:E866}, which are sensitive to the
$\Lambda$ parameters in the pion distribution functions,
but insensitive to the pion structure function at low $x$.
In the rest of this paper we focus on the analysis of the
combined set of ZEUS, H1 and E866 data.

\subsection{Combined HERA and E866 analysis}
\label{ssec:combined}

With the inclusion of the E866 $\bar d-\bar u$ asymmetry data in
the fits together with the HERA leading neutron cross sections,
the correlations between the pion flux and pion structure function
parameters decrease dramatically for all cutoff models.
For the $t$-dependent exponential model (\ref{eq:FF_exp}), for
instance, the correlations between the $\Lambda$ and $N$ parameters
are reduced to between $-8\%$ to $-16\%$ over the range of cutoffs
between $y_{\rm cut} = 0.1$ and 0.4.
The resulting $\chi^2_{\rm dof}$ profiles for all the models
are displayed in Fig.~\ref{fig:chi2}~(b).
In this case there is significantly greater discriminating
power between the form factor models, with much stronger
dependence of the fit results to the value of $y_{\rm cut}$.

In particular, the $s$-dependent exponential (\ref{eq:FF_LF}),
$k_\perp$ cutoff (\ref{eq:FF_kT2cut}) and Bishari
(\ref{eq:FF_Bishari}) models all yield large
$\chi^2_{\rm dof} \gtrsim 2$ for the entire range of
$y_{\rm cut}$ values spanned.
In fact, for the Bishari model the $\chi^2_{\rm dof}$ values
are extremely large and off the vertical scale shown in
Fig.~\ref{fig:chi2}.  This merely reflects the absence of
any $\Lambda$ dependence in the pion flux, and is consistent
with the findings in Sec.~\ref{sec:E866}.
For the $k_\perp$ cutoff model, the large $\chi^2_{\rm dof}$
values are related to the fact that a sharp cutoff does not
provide a realistic description of the data at $k_\perp \gg 0$.

\begin{figure}[t]
\includegraphics[width=16cm]{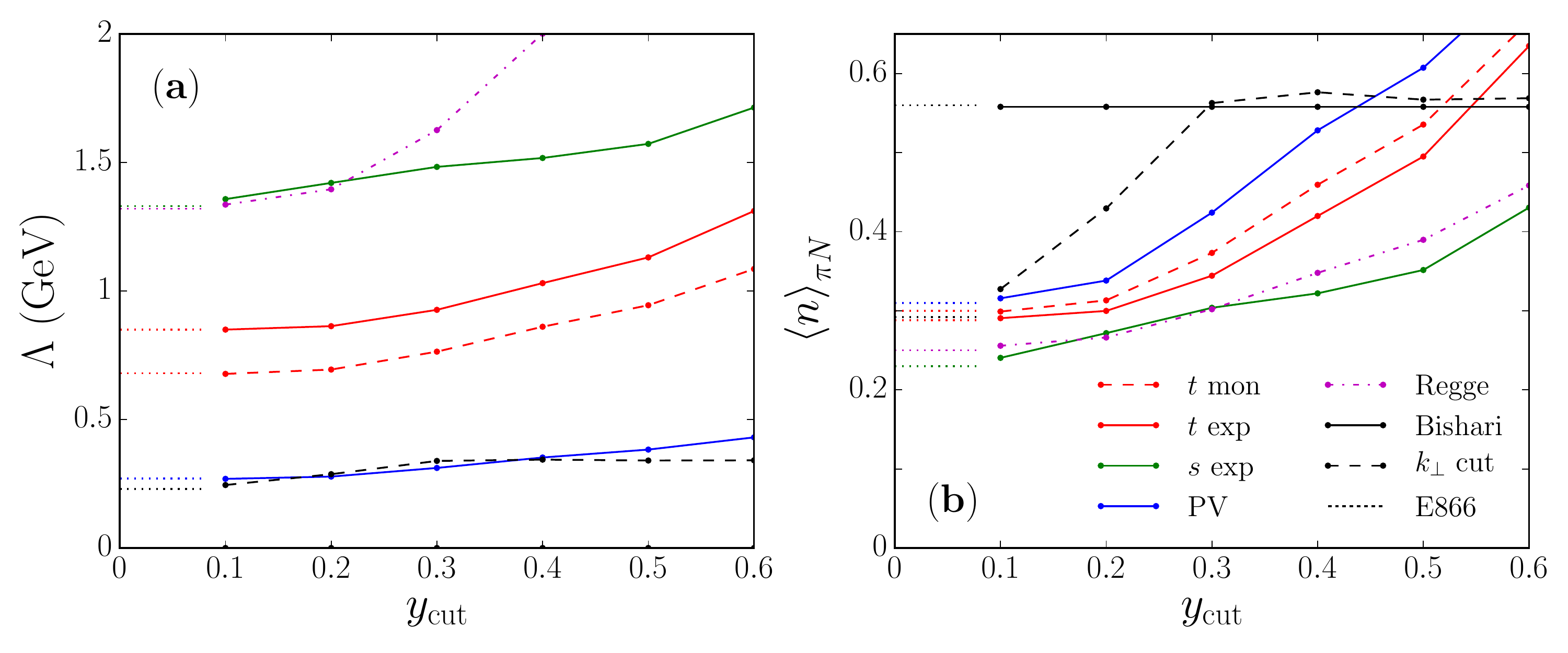}
\caption{Dependence on $y_{\rm cut}$ in the HERA data of the
	fitted values of
	{\bf (a)} the form factor cutoffs $\Lambda$ and
	{\bf (b)} the pion multiplicities $\langle n \rangle_{\pi N}$
	for various cutoff models, for the combined HERA and
	E866 fit.  The dotted horizontal extensions at small
	$y_{\rm cut}$ denote the results from fits to the E866
	data only.}
\label{fig:mom}
\end{figure}

Interestingly, the $s$-dependent exponential model, which gave
reasonably good $\chi^2$ values for the HERA data, has difficulty
in describing the $\bar d-\bar u$ asymmetry, as was evident in
Sec.~\ref{sec:E866} where a $\chi^2_{\rm dof} \sim 2$ was also
found for the fit to the E866 data alone (see Table~\ref{tab:E866}).
The poor fit to the small-$y$ HERA and E866 data can be attributed
to the functional form of the $s$-dependent form factor in
Eq.~(\ref{eq:FF_LF}).  In particular, at small values of $y$
the $\pi N$ invariant mass $s \sim k_\perp^2/y$ becomes
increasingly large, rendering the form factor zero in the $y \to 0$
limit even for finite $k_\perp$.  This gives rise to much stronger
suppression at low $y$, which is already visible in the shapes of
the splitting functions $f_N^{\rm (on)}$ in Fig.~\ref{fig:fy}~(a).
A similar suppression would arise for $u$-dependent form factors
(see Sec.~\ref{ssec:FFs}), since $u \sim k_\perp^2/y$ at low $y$,
if these were applied to splitting functions beyond the on-shell
contributions discussed in this work.
This suppression does not occur for the $t$-dependent form factors,
on the other hand, which depend on the variable
$t \sim -k_\perp^2/(1-y)$ at small $y$.
Through the convolution formula (\ref{eq:conv}), less strength at
small $y$ also translates into suppression of the calculated PDFs at
small $x$ values, which is also visible in Fig.~\ref{fig:dubar_E866}
for the $s$-dependent model.

For the other models (namely, $t$-dependent exponential and monopole,
Pauli-Villars, and Regge exponential), reasonable fits with
$\chi^2_{\rm dof} \lesssim 1$ are obtained for $y_{\rm cut}$
up to 0.2, and for the $t$-dependent exponential (\ref{eq:FF_exp})
[and to a lesser extent the $t$-dependent monopole (\ref{eq:FF_mon})]
model also at $y_{\rm cut}=0.3$.
For larger $y_{\rm cut}$ values the $\chi^2_{\rm dof}$ increases
rapidly, and no model is able to give an adequate description of
the combined data sets for $y_{\rm cut} \gtrsim 0.4$.

The larger $\chi^2_{\rm dof}$ values are in fact associated with
increasing cutoffs $\Lambda$, and correspondingly larger values
of the pion multiplicities $\langle n \rangle_{\pi N}$, as
Fig.~\ref{fig:mom} illustrates.  For all the models other than
Bishari (for which the pion flux is independent of $\Lambda$
and hence of $y_{\rm cut}$), the pion multiplicities for
$y_{\rm cut} \lesssim 0.2$ are similar to the values
$\langle n \rangle_{\pi N} \sim 0.3$ obtained in Sec.~\ref{sec:E866}
from the $\bar d-\bar u$ constraints alone.
For reference, the dotted horizontal lines in Fig.~\ref{fig:mom}
at low $y_{\rm cut}$ represent the values of the cutoffs and pion
multiplicities from the E866-only fits, as in Table~\ref{tab:E866}.
Recall that for too large cutoffs, or multiplicities
$\langle n \rangle_{\pi N} \gtrsim 0.5$, the probability of
multi-pion exchanges becomes non-negligible, and the justification
for restricting the calculation to one-pion exchange is more
questionable \cite{Thomas84, Speth98}.

\begin{figure}[t]
\includegraphics[width=17cm]{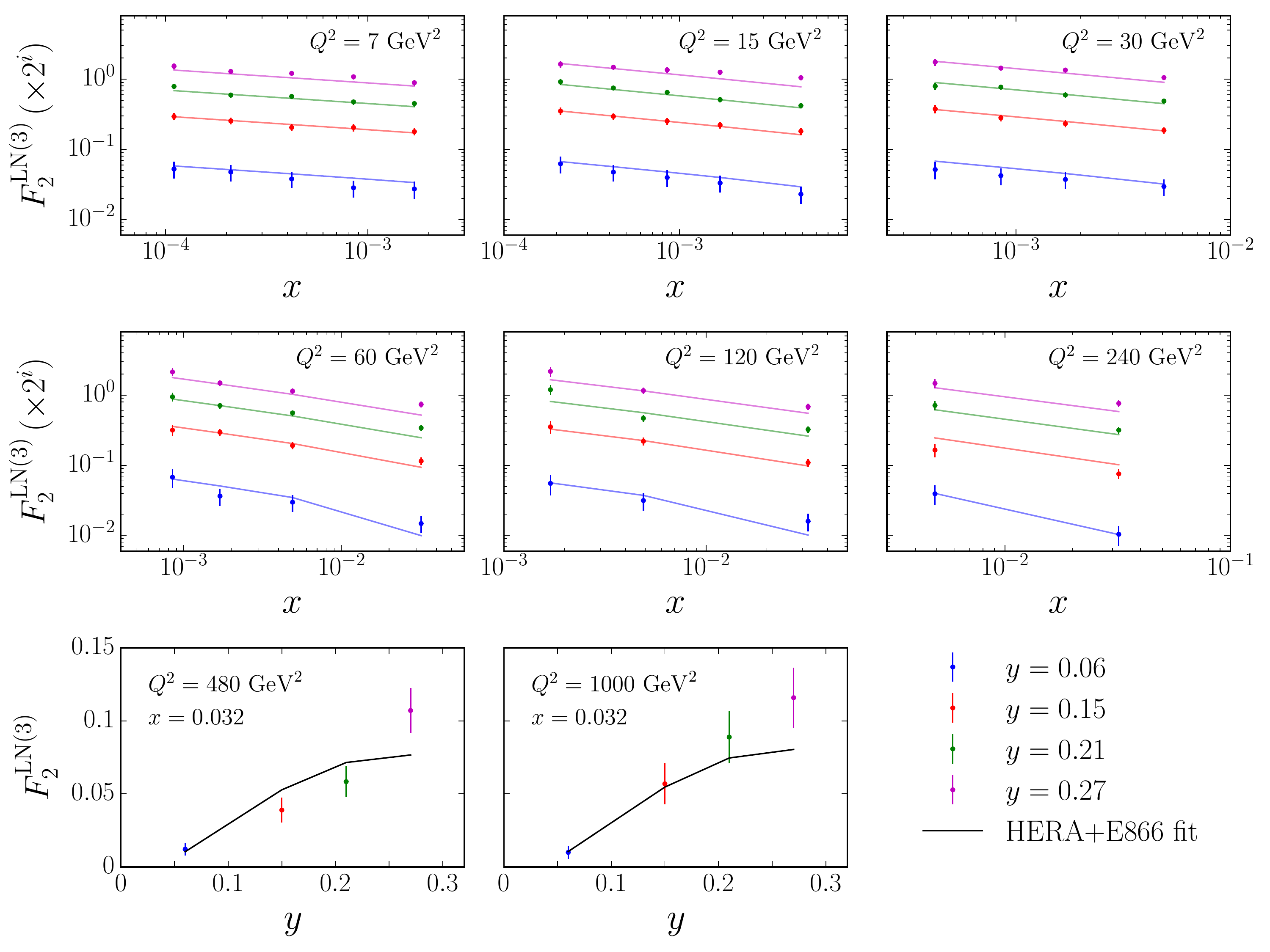}
\caption{Leading neutron structure function $F_2^{\rm LN(3)}$ from
	ZEUS \cite{ZEUS_02} as a function of $x$ at fixed values of
	$Q^2$ and $y$.  The panels at $Q^2=480$ and 1000~GeV$^2$ are
	shown as a function of $y$ for fixed $x=3.2 \times 10^{-2}$.
	The fitted results have been computed for the $t$-dependent
	exponential model (\ref{eq:FF_exp}) with $y_{\rm cut}=0.3$.
	For clarity, the values of $F_2^{\rm LN(3)}$ in the first
	six panels (for $Q^2 \leqslant 240$~GeV$^2$) have been offset
	by multuplying by a factor $2^i$ for $i=0$ (for $y=0.06$)
	to $i=3$ (for $y=0.27$).}
\label{fig:zeus}
\end{figure}

\begin{figure}[t]
\includegraphics[width=17cm]{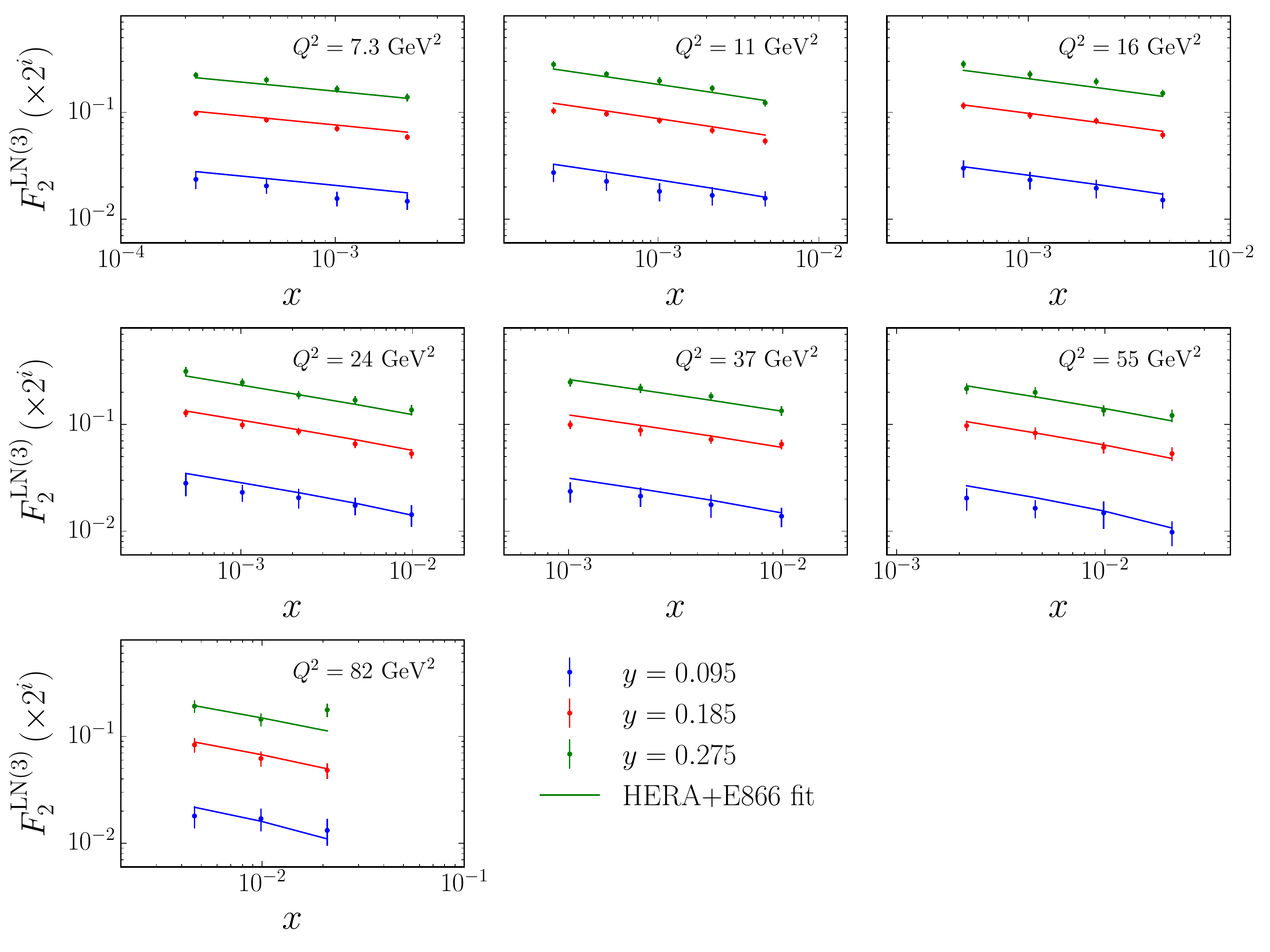}
\caption{Leading neutron structure function $F_2^{\rm LN(3)}$ from
	H1 \cite{H1_10} as a function of $x$ at fixed values of
	$Q^2$ and $y$.  The fitted results have been computed for
	the $t$-dependent exponential model (\ref{eq:FF_exp}) of
	the pion flux with $y_{\rm cut}=0.3$.
	For clarity, the values of $F_2^{\rm LN(3)}$ have been offset
	by multuplying by a factor $2^i$ for $i=0$ (for $y=0.095$)
	to $i=3$ (for $y=0.275$).}
\label{fig:h1}
\end{figure}

Reasonable values of $\langle n \rangle_{\pi N}$ are still obtained,
however, for $y_{\rm cut} = 0.3$ for the $t$-dependent exponential
and monopole, $s$-dependent and Regge exponential models, although
with the exception of the $t$-dependent exponential model all of these
give somewhat larger $\chi^2_{\rm dof} \gtrsim 2$.
Taken together, the results for the $\chi^2_{\rm dof}$, $\Lambda$
and $\langle n \rangle_{\pi N}$ profiles point to the $t$-dependent
exponential model (\ref{eq:FF_exp}) as the one best able to account
for the combined ZEUS and H1 leading neutron data and the E866
$\bar d-\bar u$ asymmetry over the largest range of $y$.

Taking the $t$-dependent exponential model with $y_{\rm cut}=0.3$
as the optimal result of our fits, in Figs.~\ref{fig:zeus} and
\ref{fig:h1} we show the spectra of leading neutrons from the
ZEUS \cite{ZEUS_02} and H1 \cite{H1_10} experiments, respectively.
For the ZEUS data we convert the measured ratios $r$ in
Eq.~(\ref{eq:r_def}) to an absolute cross section by multiplying
the ratio by the inclusive proton $F_2^p$ structure function,
Eq.~(\ref{eq:r}).
The resulting structure function $F_2^{\rm LN(3)}$ in
Fig.~\ref{fig:zeus} is plotted as a function of $x$ at fixed $Q^2$
values from $Q^2=7$ to 1000~GeV$^2$, for individual $y$ bins at
average values of $y = 0.06, 0.15, 0.21$ and 0.27.
Because the highest two $Q^2$ bins at $Q^2=480$ and 1000~GeV$^2$
contain only one $x$ value, $x=3.2 \times 10^{-4}$, we combine
these data to show the structure function as a function of $y$.
The comparison in Fig.~\ref{fig:zeus} between the data and the
fitted results shows very good agreement across all kinematics,
with the slopes in $x$ and $y$ well reproduced.
The errors on the data points shown include statistical and
systematic uncertainties added in quadrature, including an
acceptance uncertainty of $\sim 5\%$ and a normalization error
of 4\%.  For the lowest-$y$ data points at $y=0.06$ there is a
large, $\sim 25\%$ systematic uncertainty from the energy scale
uncertainty, which inflates the overall error at these points
relative to the data at larger $y$.
Uncertainties from the parametrization of the inclusive $F_2^p$
structure function are smaller than the experimental errors on
$r$ and are not included.

\begin{table}[tb]
\begin{center}
\caption{Fit parameters from the combined ZEUS, H1 and E866 fit for
	the cutoff $\Lambda$ and pion structure function parameters
	$N$, $a_0$ and $a_1$ for several fits: our optimal fit for
	the $t$-dependent exponential model (\ref{eq:FF_exp}) with
	$y_{\rm cut}=0.3$ (shown in boldface), and comparable fits
	with $y_{\rm cut}=0.2$ for the $t$-dependent exponential and
	monopole models.
	For reference the corresponding values of the pion
	multiplicities $\langle n \rangle_{\pi N}$ are also given,
	as are the number of fitted points and $\chi^2_{\rm dof}$. \\}
\begin{tabular}{c|c|c|c}			\hline
~model~
	&~$\bm{t}$~{\bf exp} $\bm{(y_{\rm cut}=0.3)}$~
	&~$t$ exp ($y_{\rm cut}=0.2$)~
	&~$t$ mon ($y_{\rm cut}=0.2$)~		\\ \hline
$\Lambda$~(GeV)~
	& $\bm{0.927 \pm 0.003}$
	& $0.863 \pm 0.004$
	& $0.694 \pm 0.005$ 			\\
$\langle n \rangle_{\pi N}$
	& {\bf 0.34}
	& 0.30
	& 0.31 					\\
$N$
	& $\bm{0.084 \pm 0.009}$
	& $0.083 \pm 0.016$
	& $0.091 \pm 0.016$			\\
$a_0$
	& $\bm{-0.0033 \pm 0.0123}$
	& $-0.0074 \pm 0.0207$
	& $-0.0047 \pm 0.0208$			\\
$a_1$
	& $\bm{-0.257 \pm 0.015}$
	& $-0.247 \pm 0.016$
	& $-0.253 \pm 0.014$			\\ \hline
$\chi^2_{\rm dof}$
	& {\bf 1.27}
	& 0.65
	& 0.80					\\
\# data points~
	& {\bf 202}
	& 123
	& 123					\\ \hline
\end{tabular}
\label{tab:combined}
\end{center}
\end{table}

\begin{figure}[t]
\includegraphics[width=12cm]{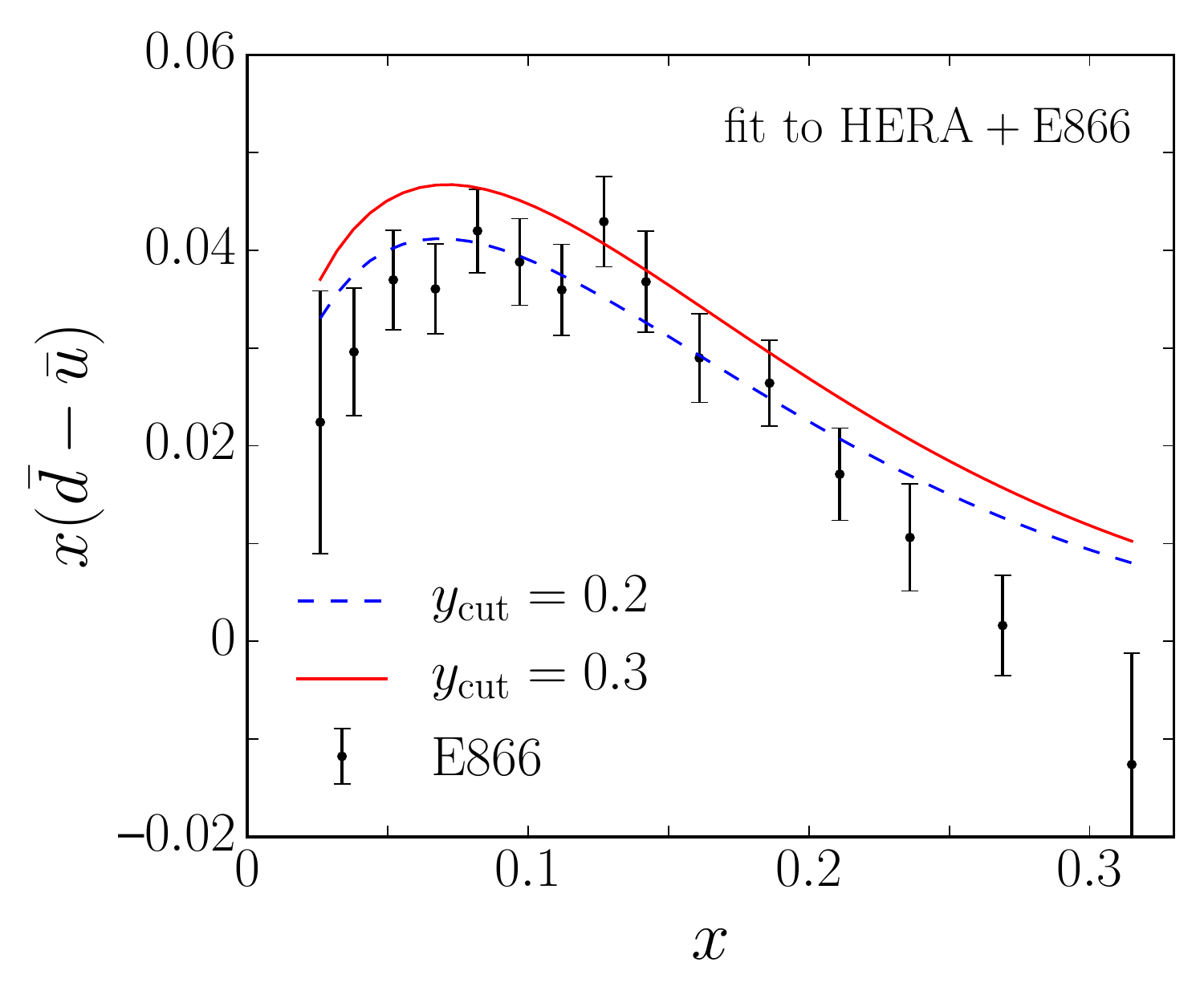}
\caption{Flavor asymmetry $x(\bar d-\bar u)$ from the combined
	fit to the HERA leading neutron \cite{ZEUS_02, H1_10}
	and E866 Drell-Yan \cite{E866} data, for cuts on the
	HERA data of $y_{\rm cut}=0.2$ (blue dashed curve)
	and 0.3 (solid red curve).}
\label{fig:dubar_final}
\end{figure}

Similarly good agreement with the measured leading neutron spectra is
obtained for the H1 data \cite{H1_10} at average $y=0.095, 0.185$ and
0.275 in Fig.~\ref{fig:h1}, in which the absolute $F_2^{\rm LN(3)}$
structure function was obtained directly over a range of $Q^2$
between 7.3 and 82~GeV$^2$.
The H1 leading neutron data were collected during the 2006--2007 run
and represent an integrated luminosity of 122~pb$^{-1}$, or about
3 times that of the ZEUS data in the DIS region.
Consequently, the statistical uncertainties of the H1 data are
smaller than those for the ZEUS leading neutron spectra.
Note that in the calculations of the leading neutron structure
functions the appropriate transverse momentum acceptance cuts of
$k_\perp^2 < 0.43\, (1-y)^2$~GeV$^2$ and $k_\perp^2 < 0.04$~GeV$^2$
were applied for the ZEUS and H1 data, respectively.

For completeness, we list in Table~\ref{tab:combined} the parameters
fitted in the combined analysis, namely the cutoffs $\Lambda$ and
$F_2^\pi$ parameters $N$, $a_0$ and $a_1$, for our optimal fit, the
$t$-dependent exponential model of the pion flux with $y_{\rm cut}=0.3$.
For comparison we also list the parameters for two comparable fits,
for the $t$-dependent exponential and $t$-dependent monopole models
with $y_{\rm cut}=0.2$.  Also listed for reference are the pion
multiplicities corresponding to the $\Lambda$ values and the
$\chi^2_{\rm dof}$ for the fits.

For our optimal model from the fit to the combined HERA + E866 data
sets, as a consistency check we show in Fig.~\ref{fig:dubar_final} the
resulting $\bar d - \bar u$ asymmetry compared with the E866 data.
The quality of the fit is similar to the fit to the E866 data alone
in Sec.~\ref{sec:E866}, as is also indicated by the similar values
for the cutoffs $\Lambda$ in Tables~\ref{tab:E866} and
\ref{tab:combined}.  For comparison we also plot the results of the
fit with the $t$-dependent exponential model for $y_{\rm cut}=0.2$,
which gives a similar cutoff to that in the E866-only fit in
Table~\ref{tab:E866} ($\Lambda = 0.85$~GeV) and hence a slightly
better fit to the E866 data.
Overall, the comparison in Fig.~\ref{fig:dubar_final} clearly
demonstrates the consistency of the one-pion exchange description,
and in particular the model of the pion flux with the $t$-dependent
exponential form factor (\ref{eq:FF_exp}), of both the HERA
leading neutron cross sections and the $\bar d-\bar u$ asymmetry.

\subsection{Pion structure function at small $x$}
\label{ssec:F2pi}

\begin{figure}[t]
\includegraphics[width=16.5cm]{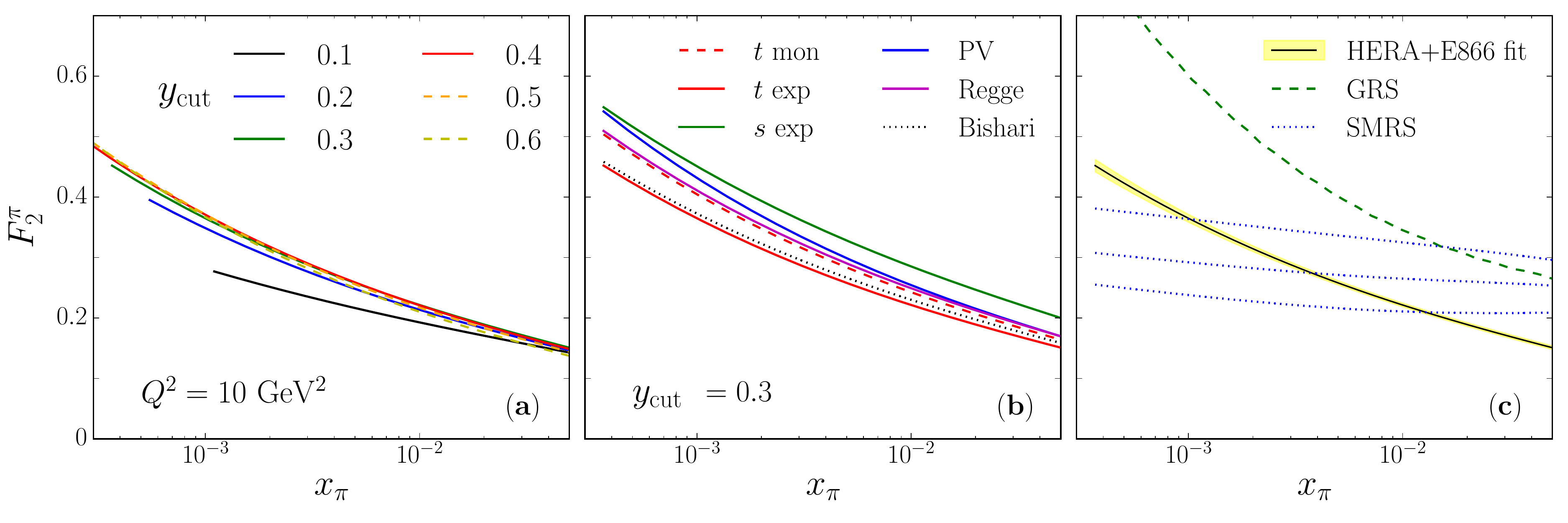}
\caption{Pion structure function $F_2^\pi$ as a function of $x_\pi$
	at $Q^2=10$~GeV$^2$, extracted from a simultaneous fit to
	the ZEUS and H1 leading neutron data and the E866
	$\bar d-\bar u$ asymmetry for
	 {\bf (a)} the $t$-dependent exponential model with
	different $y_{\rm cut}$ values,
	 {\bf (b)} various models at fixed $y_{\rm cut}$, and
	 {\bf (c)} the best fit $t$-dependent exponential model with
	$y_{\rm cut}=0.3$, compared with the GRS \cite{GRS} and
	SMRS \cite{SMRS} parametrizations, with the latter shown
	for a 10\% (lowest), 15\% (central) and 20\% (highest)
	pion sea.}
\label{fig:F2pi}
\end{figure}

Having systematically quantified the efficacy of the various
pion exchange models in describing the HERA leading neutron and
E866 $\bar d-\bar u$ asymmetry data, we can now assess whether
and to what extent the combined analysis is able to unambiguously 
determine the $x_\pi$ dependence of the pion structure function.
Choosing the $t$-dependent exponential model for the $\pi NN$ form
factor (\ref{eq:FF_exp}) as the one best capable of giving a
consistent description of the data over the largest range of
kinematics, in Fig.~\ref{fig:F2pi}~(a) we illustrate the
stability of the results for $F_2^\pi$ with respect to the
value of $y_{\rm cut}$, at a fixed $Q^2=10$~GeV$^2$.
With the exception of the $y_{\rm cut}=0.1$ fit, the extracted
$F_2^\pi$ shows remarkable stability across all cuts up to the
optimal $y_{\rm cut}=0.3$ and even beyond, over the entire range of
$x_\pi \gtrsim 4 \times 10^{-4}$ constrained by the ZEUS and H1 data.
Note that each of the curves is plotted for $x_\pi$ down to different
values of $x_\pi^{\rm min} = x_{\rm min}/y_{\rm cut}$ because of the
varying $y_{\rm cut}$ values in each fit.

Although the $t$-dependent exponential model gave the smallest
$\chi^2_{\rm dof}$ of all models in the combined fit, up to
$y_{\rm cut}=0.4$ [see Fig.~\ref{fig:chi2}~(b)], the dependence
of the fitted $F_2^\pi$ on the functional form of the $\pi NN$
form factor is rather weak, as Fig.~\ref{fig:F2pi}~(b) illustrates
for $y_{\rm cut}=0.3$.  Interestingly, the best fit model gives
the smallest $F_2^\pi$ result, with the largest magnitude
(some $20\%-25\%$ larger) found for the $s$-dependent exponential
model (which also has a $\chi^2_{\rm dof} \approx 2.5$).

On the other hand, for a given model the propagated fit errors
from the analysis are rather small, as indicated by the band
around the extracted $F_2^\pi$ in Fig.~\ref{fig:F2pi}~(c) for
the $t$-dependent exponential model with $y_{\rm cut}=0.3$.
The PDF error is also generally substantially smaller than the
difference between our fitted result for $F_2^\pi$ and the values
from the SMRS \cite{SMRS} and GRS \cite{GRS} global PDF analyses,
extrapolated to the small-$x$ region of HERA kinematics.
In particular, while our fitted $F_2^\pi$ has a similar shape to
the GRS parametrization, its magnitude is $\approx 30\%-40\%$
smaller at $x_\pi \approx 10^{-3} - 10^{-2}$.
The magnitude is closer to the result from the SMRS
parametrization at similar $x_\pi$ values, but the latter
shows considerably less variation with $x_\pi$.

Since prior to the HERA leading neutron experiment there were
no data with any sensitivity to the small-$x_\pi$ region,
the SMRS fit to the $\pi N$ Drell-Yan and prompt photon data
considered three cases for the (unconstrained) pion sea,
with 10\%, 15\% and 20\% of the pion's momentum carried by
sea quarks and gluons at a scale of $Q^2=4$~GeV$^2$.
Our results for the extracted pion structure function favor
the 20\% scenario for the sea at $x_\pi \approx 10^{-3}$,
but are closer to the 10\% scenario at $x_\pi \approx 10^{-2}$.
At larger values of $x_\pi \gtrsim 10^{-2}$ our fit is less reliable,
as it does not include the $\pi N$ Drell-Yan and prompt photon
constraints on the large-$x_\pi$ region, at which our simple
parametrization of $F_2^\pi$ in Eq.~(\ref{eq:F2pi}) is no longer
expected to be adequate.
A more complete QCD-based analysis in terms of the pion valence,
sea and gluon PDFs, fitting all available data including the HERA
leading neutron spectra, would be necessary in order to describe
$F_2^\pi$ over the entire $x_\pi$ region.

\section{Constraints from future tagged DIS experiments}
\label{sec:TDIS}

The analysis in the previous sections enabled us to establish the
models and parameters which are best able to describe the existing
data sensitive to pionic degrees of freedom in the nucleon.
While the flavor asymmetry from the E866 Drell-Yan data is sensitive
to the pion distribution function in the nucleon and the pion PDFs
at large parton momentum fractions $x_\pi$ of the pion, the HERA
leading neutron data provide information on the pion PDFs at small
$x_\pi$, once constraints on the pion flux are included from the
$\bar d-\bar u$ asymmetry.
Clearly it would be helpful to have data at complementary kinematics
to those of HERA and E866, which could enable further constraints
to be placed on the pion flux and pion structure function parameters
independently.

The upcoming tagged DIS (TDIS) experiment at Jefferson Lab \cite{TDIS}
plans to take data on the production of leading protons from an
effective neutron target in the reaction $e n \to e p X$, which,
in analogy with the HERA leading neutron leptoproduction,
can be described at small $y$ through the exchange of a $\pi^-$.
In the proposed experiment, the effective neutron target will be
prepared by tagging spectator protons with momenta between 60~MeV
and 400~MeV at backward kinematics in the DIS of the electron from
a deuteron nucleus, using the same technique that was developed for
the measurement of the neutron structure function in the BONuS
experiment at Jefferson Lab \cite{BONuS}.
In this section we use the fit results from the analysis of the
HERA and E866 data in Sec.~\ref{sec:HERA} to estimate the
leading proton (LP) structure function at kinematics relevant
for the TDIS experiment.  In analogy with the neutron structure
function $F_2^{\rm LN(3)}$ in Eq.~(\ref{eq:F2LN3}), we define
the LP structure function as
\begin{eqnarray}
F_2^{\rm LP(3)}(x,Q^2,y)
&=& f_{\pi^- p}(y)\, F_2^\pi(x_\pi, Q^2),
\label{eq:F2LP3}
\end{eqnarray}
where we have used isospin symmetry to equate the $\pi^+$ and $\pi^-$
structure functions.  Isospin symmetry also implies equivalence
between the $p \to \pi^+ n$ and $n \to \pi^- p$ splitting functions,
$f_{\pi^- p}(y) = f_{\pi^+ n}(y)$ from Eq.~(\ref{eq:fpi+n})

\begin{figure}[t]
\includegraphics[width=16.5cm]{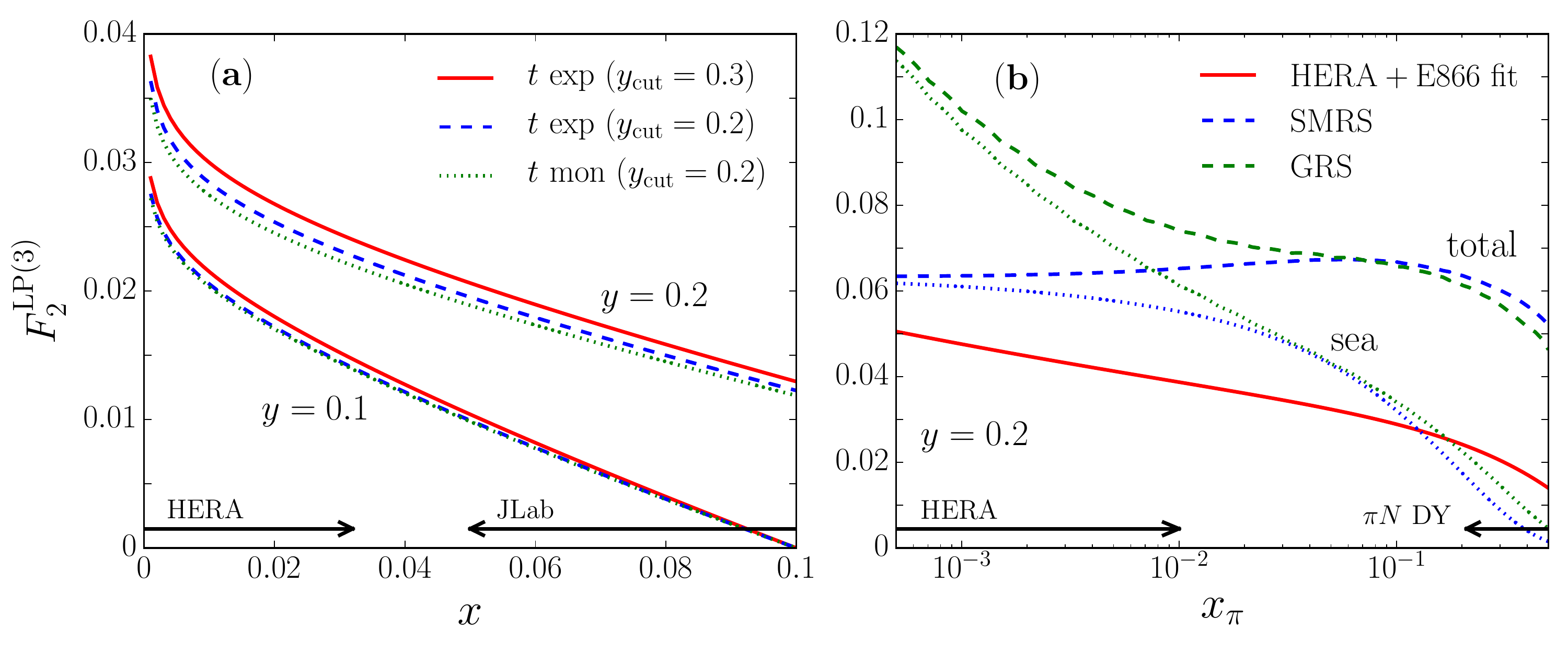}
\caption{Leading proton structure function $F_2^{\rm LP(3)}$
	in TDIS kinematics \cite{TDIS} at $Q^2=2$~GeV$^2$
	{\bf (a)}~as~a~function of $x$ at fixed $y=0.1$ and 0.2,
	for the $t$-dependent exponential model with $y_{\rm cut}=0.3$
	(red solid curves), and the $t$-dependent exponential
	(blue dashed curves) and monopole (green dotted curves)
	models with $y_{\rm cut}=0.2$, and
	{\bf (b)}~as a function of $x_\pi$ at fixed $y=0.2$ for
	the $t$-dependent exponential model with $y_{\rm cut}=0.3$
	(red solid curves), compared with the SMRS \cite{SMRS}
	(blue) and GRS \cite{GRS} (green) parametrizations for
	the total (dashed) and sea only (dotted) contributions.
	The horizontal arrows at the bottom of the panels indicate
	the reach of the HERA data at low $x$, the projected TDIS
	at Jefferson Lab data coverage at high $x$, and the region
	at large $x_\pi$ where the pion PDFs are constrained by
	the $\pi N$ Drell-Yan data \cite{E615}.}
\label{fig:TDIS}
\end{figure}

The TDIS experiment will measure the semi-inclusive $e d \to e p p X$
cross section over the kinematic range corresponding to the parton
momentum fraction in the neutron of
\mbox{$0.05 \lesssim x \lesssim 0.1$}, and for
$0.05 \lesssim y \lesssim 0.3$, at an average $Q^2 = 2$~GeV$^2$.
In Fig.~\ref{fig:TDIS}~(a) the leading proton structure function
is shown as a function of $x$ for typical TDIS kinematics,
for fixed values of $y=0.1$ and 0.2.  To illustrate the model
dependence of the results, the structure function calculated
using the parameters from the $t$-dependent exponential form
factor with $y_{\rm cut}=0.3$ is compared with those using the
$t$-dependent exponential and monopole models with $y_{\rm cut}=0.2$.
The differences between the models are relatively small over the
entire range of kinematics considered.
Note, however, that the fitted results have been extrapolated
from the region where they are constrained by the HERA data,
for which the largest $x$ is $3.2 \times 10^{-2}$, to the TDIS
kinematics where $x \gtrsim 0.05$.
Furthermore, since the lowest $Q^2$ for the HERA data is 7~GeV$^2$,
the fitted pion structure function is extrapolated to the
$Q^2=2$~GeV$^2$ value relevant for the TDIS experiment using
the functional form in Eqs.~(\ref{eq:F2pi})--(\ref{eq:eta}).
Comparing with the NLO evolution of phenomenological PDFs,
the uncertainty from our approximate $Q^2$ evolution is of the
order 20\% between $Q^2 \approx 2$~GeV$^2$ and 10~GeV$^2$.
This does not affect, however, the observation that the
dependence of $F_2^{\rm LP(3)}$ on the pion flux model is
weak at Jefferson Lab kinematics.

Plotted as a function of $x_\pi$, the leading proton structure
function $F_2^{\rm LP(3)}$ for the best fit $t$-dependent exponential
model with $y_{\rm cut}=0.3$ is shown in Fig.~\ref{fig:TDIS}~(b)
for a fixed $y=0.2$.  While the results are constrained by the
HERA data at small $x_\pi$, because of the simple choice of
parametrization for $F_2^\pi$ in Eq.~(\ref{eq:F2pi}) our
calculation is effectively an extrapolation for
$x_\pi \gtrsim 10^{-2}$.
Comparing with the $F_2^{\rm LP(3)}$ computed from pion PDF
parametrizations evolved to $Q^2=2$~GeV$^2$, our results are
smaller than those using both the SMRS \cite{SMRS} and GRS
\cite{GRS} fits, with the differences similar to those observed
in Fig.~\ref{fig:F2pi}~(c) for $F_2^\pi$.
On the other hand, the phenomenological pion PDFs \cite{SMRS, GRS}
are fitted to the $\pi N$ Drell-Yan data \cite{E615} only down to
$x_\pi \approx 0.2$, and for smaller $x_\pi \lesssim 0.1$ the
parametrizations are unconstrained.
It is interesting, however, to observe that the differences
between the $F_2^{\rm LP(3)}$ calculated using only the sea part
of the pion structure function parametrizations and our fit are
significantly reduced at $x_\pi \gtrsim 10^{-2}$.
This may reflect the restricted form (\ref{eq:F2pi}) used for
our $F_2^\pi$ parametrization, which, while suited for describing
the small-$x_\pi$ region where the sea dominates, may not be
optimal for all $x_\pi$.
A more systematic approach in the future would be to perform a
combined global PDF analysis of leading neutron and $\pi N$
Drell-Yan data in terms of pion PDFs, separating the pion
structure function into its valence and sea components.

With the TDIS data expected to cover the region $x_\pi \gtrsim 0.1$
\cite{TDIS}, this experiment offers an important opportunity to bridge
the gap between the HERA data which can constrain the pion PDFs at
low $x_\pi$ and the $\pi N$ Drell-Yan data that have been used to
determine the pion's valence quark content at $x_\pi \to 1$.
Together with the constraints from the E866 $\bar d-\bar u$ asymmetry,
the combined data sets should be able to more precisely pin down the
partonic structure of the pion over a much more extended range of
$x_\pi$.

\section{Conclusion}
\label{sec:conclusion}

Our analysis has sought to determine whether a consistent
description of the HERA leading neutron cross sections
\cite{ZEUS_02, H1_10, ZEUS_07} can be obtained within a pion
exchange framework, while simultaneously also accounting
for the $\bar d-\bar u$ asymmetry in the proton extracted
from the E866 Drell-Yan data \cite{E866}.
Previous analyses of the HERA data alone have generally drawn
somewhat negative conclusions about whether one can reliably
extract information on the pion structure function $F_2^\pi$
at small $x_\pi$, that was not subject to large uncertainties
associated with the choice of the pion flux.
Rather than relying on assumptions about specific forms for the
pion light-cone distributions in the nucleon, we have addressed the
model dependence empirically, by performing the first comprehensive
statistical analysis of the combined HERA leading neutron and
E866 data sets, for a wide range of prescriptions adopted in
the literature for regularizing the pion--nucleon amplitudes.

Our findings suggest that we can indeed describe both HERA and E866
data within a one-pion exchange framework, if the cutoff parameters
in the $\pi NN$ form factors are fitted simultaneously with the
pion structure function.  For the E866 data, we find that almost
all the models that have adjustable cutoffs are able to provide
reasonable descriptions of the $\bar d - \bar u$ asymmetry.
The exceptions are the Bishari model \cite{Bishari}, which has
parameters fixed by hadron production data in inclusive $pp$
scattering, and the sharp $k_\perp$ cutoff model, which we
consider mostly for illustration purposes.
The E866 data are also not very sensitive to the precise
large-$x_\pi$ behavior of the pion PDFs.

For the HERA leading neutron data, since background processes
other than one-pion exchange, such as the exchange of other
mesons and absorption corrections, are known to play an
increasingly important role at large pion momentum fractions $y$,
we do not attempt to model the data over the entire range of
kinematics.  Instead we perform $\chi^2$ fits to determine the
relevant region where one-pion exchange is applicable empirically.
Most of the models considered are able to give reasonable $\chi^2$
values for $y_{\rm cut} \lesssim 0.5$.  However, fitting only the
HERA data we find large correlations between the fitted pion flux
and pion structure function parameters, suggesting that it is not
possible to unambiguously extract these independently of one
another.

On the other hand, the combined fits to both the HERA and E866 data
are significantly more restrictive, with models with $t$-dependent
form factors, such as the exponential or monopole, giving the best
descriptions of the combined data sets over the largest range of
kinematics, up to $y_{\rm cut} \approx 0.3$ \cite{Kumano98, Speth98,
MST99}.  Models with $s$-dependent form factors give poor fits, with
$\chi^2_{\rm dof} \sim 2$ irrespective of the $y_{\rm cut}$ value.
For $y_{\rm cut}=0.2$, all the models with $t$-dependent form factor
and adjustable cutoffs (exponential, monopole, Pauli-Villars and
Regge exponential) give good descriptions of the combined data sets,
with reasonable values of the average pion multiplicity,
$\langle n \rangle_{\pi N} \approx 0.3$.
A slight preference is found for the $t$-dependent exponential
model, owing to the good description ($\chi^2_{\rm dof} \sim 1$)
obtained over the largest $y$ range, up to $y_{\rm cut}=0.3$.
While the restricted $y$ regions reduce the number of data points
available for the fit, cuts of $y_{\rm cut}=0.2$ and 0.3 still
provide 123 and 202 HERA data points, respectively.

For the preferred models, excellent descriptions of the ZEUS
and H1 leading neutron spectra are obtained over the entire
range $10^{-4} \lesssim x \lesssim 0.03$ and
$7 \leqslant Q^2 \leqslant 1000$~GeV$^2$ covered by the data.
For parton momentum fractions in the pion of
$4 \times 10^{-4} \lesssim x_\pi \lesssim 0.1$, the extracted
pion structure function $F_2^\pi$ for these models is rather
weakly dependent on the choice of $y_{\rm cut}$, and indeed on
the form factor model.  Compared with existing parametrizations
of pion PDFs, which are well constrained at large $x_\pi$,
the extrapolation of the GRS fit \cite{GRS} into the HERA region
overestimates our fitted results by a factor $\sim 2$, but has a
similar shape, while the SMRS fit \cite{SMRS} is closer to our
fit in magnitude, but has a shallower $x_\pi$ dependence.
Our fitted result is somewhat smaller than both the phenomenological
parametrizations at $x_\pi \approx 0.1$, which may be due to the
limitations of our simple parametric form for $F_2^\pi$, which
is constructed for the sea region, or because our fit is not
constrained at large $x_\pi$ by the $\pi N$ Drell-Yan data.

In the near future, the SeaQuest Drell-Yan experiment \cite{SeaQuest}
at Fermilab will measure the $\bar d - \bar u$ difference up to
larger values of $x$, $x \approx 0.45$, which should allow improved
constraints on the models of the pion distribution function in the
nucleon.
Beyond that, the tagged DIS experiment \cite{TDIS} at Jefferson Lab
will provide precise information on pion exchange in leading proton
production from an effective neutron target at kinematics complementary
to the range covered by the HERA and Drell-Yan measurements.
This should reduce the uncertainty in $F_2^\pi$ in the intermediate
$x_\pi$ region, $x_\pi \sim 0.1$.

One may also examine in more detail the $k_\perp$ dependence of
leading neutron (or proton) cross sections, which was studied in
some of the HERA measurements \cite{ZEUS_07} and will be explored
in the TDIS experiment.  Comparison of the unintegrated pion flux
with the empirical transverse momentum distributions could provide
a more incisive test of the momentum dependence of the $\pi NN$
form factor.
In the longer term, a necessary goal would be to perform a global
PDF fit, in terms of both sea and valence quark PDFs, to the $\pi N$
Drell-Yan data at moderate and high values of $x_\pi$, together with
HERA leading neutron data at small $x_\pi$, and the new TDIS data on
leading proton production in the intermediate $x_\pi$ region.
We look forward to these endeavors revealing much more consisely
and completely the partonic structure of the pion, and the role of
the pion cloud in the structure of the nucleon.

\acknowledgments

We are grateful to T.~J.~Hobbs for discussions and collaboration on
portions of this work.
This work was supported by the DOE contract No.~DE-AC05-06OR23177,
under which Jefferson Science Associates, LLC operates Jefferson Lab,
DOE Contract No.~DE-FG02-03ER41260, the DOE Science Undergraduate
Laboratory Internship (SULI) Program, and the JSA Initiatives Fund.

\appendix 
\section{Hessian error analysis}
\label{sec:Hessian}

The fits to the experimental data in our analysis are performed
using standard $\chi^2$ minimization to find the optimal set of
fit parameters.  To estimate the uncertainties in the parameters,
or in various derived quantities from the model, we employ the
Hessian error technique.  The method is valid for any number of
parameters in a given model (including for the case of only one
parameter).

The essential idea of the method is to find a set of directions
in parameter space around the best fit values ($\bm{p}_0$) which
can be treated as statistically independent.  These are found by
diagonalizing the Hessian matrix $H$, whose elements are defined as
\begin{align}
H_{ij}
= \left. \frac{1}{2}
	 \frac{\partial\chi^2(\bm{p})}{\partial p_i\partial p_j}
  \right|_{\bm{p}=\bm{p_0}},
\end{align}
with $i$ ranging from 1 to the number of parameters.
The statistically independent directions (or eigendirections) of
the Hessian $H$ are labeled by $\hat{\bm{e}}_i$ and parametrize
the shifts in the parameter space,
\begin{align}  
\Delta\bm{p} = \bm{p}-\bm{p}_0 = \sum_i \xi_i \hat{\bm{e}}_i.
\end{align}  
The basic assumption in the Hessian analysis is that the probability
distribution ${\cal P}$ of the parameters $\bm{p}$ factorizes along
each eigendirection,
\begin{align}  
\mathcal{P}(\Delta\bm{p})
\cong \prod_i \mathcal{P}(\xi_i \hat{\bm{e}}_i),
\end{align}
where
\begin{align}
\mathcal{P}(\xi_i \hat{\bm{e}}_i)
= \mathcal{N}
  \exp\left[- \frac{1}{2} \chi^2 (\bm{p}_0 + \xi_i\hat{\bm{e}}_i)
      \right],
\end{align}
and $\mathcal{N}$ is a normalization constant.
One can then perform the error propagation for a given observable
$\mathcal{O}$ along each eigendirection, and add the independent
errors in quadrature.
The errors along each individual eigendirection are given by
\begin{align}
\delta_i\mathcal{O}
= \mathcal{O}(\bm{p}+\xi^{\text{CL}}_i\hat{\bm{e}}_i)
- \mathcal{O}(\bm{p}_0),
\end{align}
where, for a given confidence level (CL), $\xi_i^{\text{CL}}$
is the boundary such that the region
$-\xi_i^{\text{CL}} \leqslant \xi_i \leqslant \xi_i^{\text{CL}}$
is the corresponding CL region for the probability distribution
$\mathcal{P}(\xi_i\hat{\bm{e}}_i)$.
The total combined uncertainty for the observable is given by
\begin{align}
\delta\mathcal{O}
= \sqrt{ \sum_i (\delta_i\mathcal{O})^2 }.
\label{eq:master}
\end{align}
If the $\chi^2$ along each eigendirection behaves quadratically
as a function of $\xi_i$, then setting $\xi_i^{\text{CL}}=1$
induces a change in the $\chi^2$ by one unit.  This occurs, for
example, if the model is linear in the parameters, for which
Gaussian behavior holds.  However, for parameters that are weakly
constrained by the data one does not observe Gaussian behavior.

This method avoids the use of a $\Delta\chi^2$ criterion, which is
sometimes used in the literature for inflating errors when fitting
to incompatible data sets.  Moreover, the $\chi^2$ can be treated
as an observable, and its shift for a given CL can be quantified
using Eq.~(\ref{eq:master}).  In particular, by setting the CL
equal to 1$\sigma$ we can asses whether the errors satisfy Gaussian
statistics.

\section{Likelihood analysis}
\label{sec:t-statistics}

In this section we describe our statistical method for comparing
the degree of compatibility (DOC) among models.  The method is based
on hypothesis testing using the standard t-statistic, $\tau$, defined
as the log-likelihood ratio
\begin{align}
\tau^{D}_{AB}
= 2\ln \frac{\mathcal{L}(D|\mathcal{M}_B)}
	    {\mathcal{L}(D|\mathcal{M}_A)},
\label{eq:t-statistic}
\end{align}
where $\mathcal{L}$ is the likelihood function, $D$ represents the
data, and the models $\mathcal{M}_A$ and $\mathcal{M}_B$ are the
null and alternative hypotheses, respectively.  For a given model
$\mathcal{M}$, the likelihood function is proportional to
\begin{align}
\mathcal{L}(D|\mathcal{M})
\propto \prod_i
  \exp \left[-\frac{1}{2}
 	     \left(\frac{D_i-T_i(\mathcal{M})}{\delta D_i}
	     \right)^2
       \right],
\end{align}
where $T_i(\mathcal{M})$ are theory predictions for the observable
$D_i$ with uncertainty $\delta D_i$ in a given kinematic bin $i$.
Using this definition, one can construct the probability distribution
$\mathcal{P}_{\chi}(\tau)$ from a sample of $\tau$ values computed
from Eq.~(\ref{eq:t-statistic}) using pseudodata sets $\{D\}_{\chi}$
generated from a given model $\mathcal{M}_{\chi}$.
This is achieved by drawing each data point in the data set
from a normal distribution $\mathcal{N}(\mu,\sigma)$, with
$\mu = T_i(\mathcal{M}_{\chi})$ and $\sigma$ equal to the
experimental uncertainties $\delta D_i$.

The DOC between any two models can then be expressed in terms of
the overlapping area between their corresponding t-distributions.
In particular the DOC between the models $\mathcal{M}_A$ and
$\mathcal{M}_B$ is given by
\begin{align}
{\rm DOC}(A,B)
= \int_{-\infty}^{\infty} d\tau~
  {\rm min}\left[\mathcal{P}_A(\tau),\mathcal{P}_B(\tau)\right].
\label{eq:DOC}
\end{align}
A compatibility of 100\% indicates that the models cannot be
distinguished by the data.

In our current analysis we select the null and alternative hypotheses
to be the models that have the best and worst description of the data
(using the minimun $\chi^2$ as a criterion), respectively, and the
DOC is computed with respect to the best model.


\end{document}